# Grain Alignment: Role of Radiative Torques and Paramagnetic Relaxation

A. Lazarian (*University of Wisconsin-Madison, USA*), B-G Andersson (*SOFIA Science Center, USA*), Thiem Hoang (*Ruhr University Bochum and Goethe University Frankfurt, Germany; Current address: Canadian Institute for Theoretical Astrophysics, 60 St George St., Toronto, Canada*)

**1 Introduction**

Polarization arising from aligned dust grains presents a unique opportunity to study magnetic fields in the diffuse interstellar medium and molecular clouds. Polarization from circumstellar regions, accretion disks and comet atmospheres can also be related to aligned dust. However, the interpretation of the polarimetric information depends crucially on our ability to quantitatively describe grain alignment. For instance, the "measurement" of magnetic fields with interstellar polarization cannot be definitive unless a quantitative and well-tested grain alignment theory is used. At the same time, the alignment of "classical" interstellar grains, i.e. in the range of $5\times10^{-6}$cm to $5\times10^{-5}$cm, has proven to be a very difficult problem and it is only now that we have a predictive theory that can be observationally tested.

With big investments in instruments with polarimetric capabilities, e.g. for emission polarimetry like BLASTPol, PILOT, SCUBA2Pol (Bastien et al. 2005), all sky polarimetry with PLANCK, advances in extinction polarimetry with the Galactic Plane Infrared Polarization Survey (GPIPS) that allows measurements of thousands of sky fields, it is essential to have an adequate grain alignment theory to be able to interpret the wealth of polarimetry data in terms of magnetic fields and variations of other astrophysical parameters.

Grain alignment has a reputation of being a very tough astrophysical problem of very long standing. Indeed, for a long time after the discovery of dust-induced starlight dichroic polarization in 1949 (Hall 1949, Hiltner 1949) the mechanism of alignment was both enigmatic and illusive. Works by great minds like Lyman Spitzer and Edward Purcell moved the field forward in terms of understanding the basic physics of grain dynamics, but, nevertheless, grain alignment theory did not have predictive powers for a long time.

An extended discussion of the history of the grain alignment subject can be found in the review by Lazarian (2003). Below we briefly mention several milestones in our understanding of grain alignment.

Several mechanisms were proposed and elaborated on, to various degrees, to account for the grain alignment (see Lazarian 2007), including the "textbook solution" of paramagnetic alignment by Davis and Greenstein (1951), which matured through intensive work after its introduction (e.g. Jones and Spitzer 1967, Purcell 1979, Spitzer and McGlynn 1979, Mathis 1986, Roberge et al. 1993, Lazarian 1997, Roberge and Lazarian 1999). However, paramagnetic alignment fails as the major alignment mechanism for the aforementioned "classical" grains that are used for magnetic field tracing (see Section 6.4). It is likely, however, that it can align grains smaller than $10^{-6}$ cm, e.g. Polycyclic Aromatic Hydrocarbon (PAH) grains (Lazarian and Draine 2000, Hoang et al. 2013).

Mechanical stochastic alignment was pioneered by Gold (1952a, b), who concluded that supersonic flows could align grains rotating thermally. Further advancement of the mechanical alignment mechanism (e.g. Lazarian 1994, 1995a) allowed one to extend the range of its applicability, but nevertheless left it as an auxiliary process. Mechanical alignment of helical grains discussed in Lazarian (1995b), Lazarian, et al. (1997), and Lazarian and Hoang (2007b) is similar to the radiative torque alignment that this review mostly deals with. This mechanism may seem to be more promising as it can align grains within subsonic flows. However, our preliminary Monte-Carlo simulations show that irregular grains may not show well-defined helicity (see below) in the process of grain interactions with gaseous flows and this limits the applicability of the mechanism.

In this situation the mechanism based specifically on radiative torques is the most promising (see Andersson and Potter 2007, Lazarian 2007, Whittet et al. 2008). Thus we here focus our present contribution on this mechanism. The reader is referred to Lazarian (2003) for a review of other possible alignment mechanisms.

The effect of grain alignment induced by radiative torques was discovered by Dolginov and Mytrophanov (1976). They considered a grain that exhibited a difference in the cross-section for right-handed and left-handed photons. They noticed that the scattering of unpolarized light by such a grain resulted in its spin-up. However, they could not quantify the effect and therefore their pioneering work was mostly neglected for the next 20 years. We owe the explosion of interest in, and progress on, radiative torques to Bruce Draine, who realized that the torques could be treated quantitatively with a modified version of the discrete dipole approximation code by Draine and Flatau (1994) known as DDSCAT. Empirical studies in Draine (1996), Draine and Weingartner (1996, 1997) [1], Weingartner and Draine (2003) demonstrated that the magnitude of torques is substantial for irregular shapes. Later, the spin-up of grains by radiative torques was demonstrated in laboratory conditions (Abbas et al. 2004).

It should be stressed, however, that reliable predictions for the alignment degree cannot be obtained with this "brute force" numerical approach. Indeed, radiative torques depend on many parameters, e.g. grain shape, size, composition, radiation wavelength, the angle between the radiation direction and the magnetic field. It is not practical to do numerical calculations for this vast multidimensional parameter space.

The quantitative stage of radiative torque studies required generalized theoretical models describing radiative torques. In Lazarian and Hoang (2007a) (hereafter LH07a) we proposed a simple model of radiative torques, which allowed a good analytical description of the radiative torque alignment. Grain helicity was quantified for irregular grains and identified as the cause of the grain alignment with the long grain axes perpendicular to magnetic field. This study brought the radiative torque alignment to a new quantitative stage.

The LH07a model opened new avenues for theoretical studies by including additional physical effects. In particular, the LH07a theory was elaborated and extended in the follow up studies by Lazarian and Hoang (2008; hereafter LH08) and Hoang and Lazarian (2008, hereafter HL08; 2009a,b, hereafter HL09a and HL09b). Following LH07a, we shall hereafter abbreviate RAdiative Torques as RATs.

Although the RAT mechanism dominates for interstellar grains larger than $10^{-5}$ cm, RATs are not capable of aligning grains that are substantially smaller. This arises from the fast drop of RAT efficiencies when grains are smaller than the radiation wavelength. For such small grains, other mechanisms, e.g. the traditional paramagnetic alignment (Davis and Greenstein 1951, Roberge and Lazarian 1999) and its more sophisticated modifications, e.g. the resonance paramagnetic relaxation (Lazarian and Draine 2000), become important. Finally, the observational testing of grain alignment theory is becoming essential as the theory gets quantitative and predictive.

The material of this chapter is complementary to that in ubiquitous grain alignment. In particular, our chapter is relevant to the material in chapters on polarization from high mass stars (III.12,), comets (IV.22), interstellar media (III.9), disks (III.15) and interplanetary dust (IV.24).

---

[1] Some conclusions of these pioneering studies, as we discuss later in the review, were not supported by further research. In particular, still well entrenched in the community is the idea that RATs spin up the grains, while paramagnetic dissipation does the alignment, as well as the notion of cyclic grain phase trajectories, are fallacies.

In Section 6.2 we discuss the LH07A theory of grain alignment and show how this theory is being augmented by adding additional physical effects in Sections 6.3, 6.4, and 6.5. We consider predictions of the RAT theory in Section 6.6. While our review deals mostly with RAT alignment, we consider paramagnetic alignment to address the issue of the alignment of small grains for which the RAT is inefficient (Section 6.7). The observational testing of RATs is considered in Section 6.8. A discussion of the current understanding of grain alignment is presented in Section 6.9.

## 2 RAT quantitative theory: analytical model for radiative torques

### 2.1 Introducing the model for RATs

As discussed above, a quantitative theory of RAT alignment long remained elusive after Dolginov and Mytrophanov (1976) introduced the process[2]. Draine and Weingartner (1996) used DDSCAT to calculate the radiative torques on model grains and demonstrated that these torques can dominate other torques acting on the grains. Within this study, radiative torques were acting to spin up grains while the alignment was produced by paramagnetic dissipation. This assumption is valid if the radiation field is isotropic, while it was demonstrated later in Draine and Weingartner (1997) that anisotropic radiation results in much stronger torques that themselves align grains in a process somewhat similar to one discussed in Dolginov & Mytrophanov (1976). However, their approach treated the dynamics for grains with small angular momentum not adequately. Such moments, called "crossovers" (Spitzer and McGlynn 1979), are associated with the grain changing its rotation. With the dynamics of crossovers not included, Draine and Weingartner (1997) obtained a distorted dynamics of the alignment, e.g. induced cyclic trajectories of grains. A discussion of crossovers was later presented in Weingartner and Draine (2003)[3]. Nevertheless, these important studies did not provide a physical explanation of the properties of RAT alignment, e.g. its universality. In particular, the tendency of grains to align *perpendicular* to the magnetic field lines, in the presence of RATs, remained mysterious.

One of us (AL) recalls Lyman Spitzer being excited by the prospects of RAT alignment, and insisting that if the RATs were the major mechanism, they must be a natural explanation of observations showing that grains are universally aligned with their longer axes perpendicular to magnetic field. The latter seemed unexpected in view of the high variation of properties of torques calculated with DDSCAT for different grains. Lyman Spitzer wanted a property of grains to be identified that would explain this behavior.

Helicity was identified by LH07a as the property that drives RAT alignment. A grain of right-handed helicity is defined so that, when the maximal inertia axis is parallel to the radiation beam, it can rotate in a clockwise sense around the radiation beam; a grain has left-handed helicity if it rotates anticlockwise around the radiation beam. This property was also discussed in the prophetic work of Dolginov and Mytrophanov (1976). However, they came to an erroneous conclusion that the RAT alignment should be similar to the mechanical alignment with the direction of alignment depending on the angle between the magnetic field and direction to the radiation source. In this situation, RAT alignment could not explain interstellar polarization and could be only an auxiliary alignment process. This was corrected by LH07a who adopted a model of a helical grain and calculated its dynamics.

The challenge that existed in describing RATs is that no analytical theory of irregular grain scattering existed, while the model grains of connected twisted ellipsoids adopted by

---

[2] Our calculations showed that their analytical expression for the torques obtained for a grain consisting of two ellipsoids is, in fact, incorrect

[3] As we discuss later, we disagree with the conclusion in that work that most grains would rotate subthermally and therefore necessarily should have low degree of alignment (see Lazarian & Hoang 2007a).

Dolginov and Mytrophanov (1976) and calculated analytically were not exhibiting RATs. Therefore to capture the essence of RAT alignment LH07a adopted a simple model of a grain that interacts with the radiation by simply reflecting radiation by a weightless mirror attached at an angle to its oblate body (see Fig. 1). This model represented grain helicity in terms of its interaction with the radiation flow, but did not attempt to represent the complex process of scattering light by an irregular grain. Nevertheless, LH07a showed that this simple model of a grain accurately reproduces the essential, basic, properties of RATs, as derived from numerical DDSCAT modeling. LH07a termed the model AMO, (Analytical Model). Changing the angle at which the mirror is attached, one can make both 'left-handed' and 'right-handed' model grains. For the model grain in Fig. 1 to become 'right handed' the mirror should be turned by 90°. Our studies with DDSCAT confirmed that actual irregular grains also vary in handedness and this explains the substantial differences in radiative torques on different grains reported earlier (see Draine and Weingartner 1997).

DDSCAT calculations show that different grains have different ratios of torque components, however, AMO reproduces the functional dependence of torques well. Using the functional dependence from AMO for two components of torques and adjusting their amplitude ratio, LH07a could reproduce the alignment of irregular grains of different shapes calculated numerically. With the introduction of AMO, RAT alignment stopped being mysterious, and transitioned into a predictive theory.

*2.2 AMO: description*

The model is based on an oblate spheroidal grain with an axis of maximum moment of inertia $a_1$ and minor axes $a_2$ and $a_3$. To obtain scattering properties reproducing those of a helical grain, a weightless mirror is connected to the spheroid by a rod of length $l_1$ parallel to $a_3$. The orientation of the mirror is determined by its normal vector **N**, which is assumed to be fixed in the plane $a_1a_2$ and makes an angle $\alpha$ with $a_2$. During grain rotation, the angle $\alpha$ is constant. The mirror size $l_2$ is much smaller than its distance to the ellipsoid $l_1$ (see Fig. 1a).

Physically, when photons from the incident radiation field are reflected by the mirror and spheroid, they transfer part of their momenta to the grain, which results in an instantaneous radiative torque. After one rotation period around $a_1$, the radiative torques due to scattering by the spheroid is averaged to zero due to its symmetry, while the torques due to the mirror is not averaged out.

To make a proper comparison with numerical calculations, LH07a chose a scattering system of reference, $e_1e_2e_3$ (Fig. 1b), where $e_1$ is the direction of the incident radiation **k**, $e_2$ lies in the plane of **k** and $a_1$, and $e_3 = e_1 \times e_2$. The RAT components, $Q_{ei}$, along the axes $e_i$, with i=1, 2, 3 depend on the three angles $\Theta$, $\Phi$, and $\beta$. Due to the fast variation of $\beta$, the components $Q_{ei}$ are however averaged over the range $\beta=0$ to $2\pi$.

*2.3 LH07a study: main results*

The torques obtained analytically, using an assumption of geometric optics for the model in Fig. 1a, were shown to be in good agreement with the torques calculated with DDSCAT for irregular grains (Fig. 2).

We found that for the problem of alignment only torques $Q_{e1}$ and $Q_{e2}$ mattered. The third component $Q_{e3}$ only induces grain precession, which for most situations is subdominant to the Larmor precession of the grain in the interstellar magnetic field (see Table 1 for different time scales involved). Interestingly enough, the conclusion that $Q_{e3}$ is not important in terms of the RAT alignment is also true in the presence of thermal fluctuation (see HL08) and inefficient internal relaxation (see HL09b) when the alignment of angular momentum with the axis of maximum moment of inertia ($a_1$) is not enforced.[4]

---

4 It was shown in LH07a that the only component of RATs present for an ellipsoidal grain is $Q_{e3}$. Naturally, this

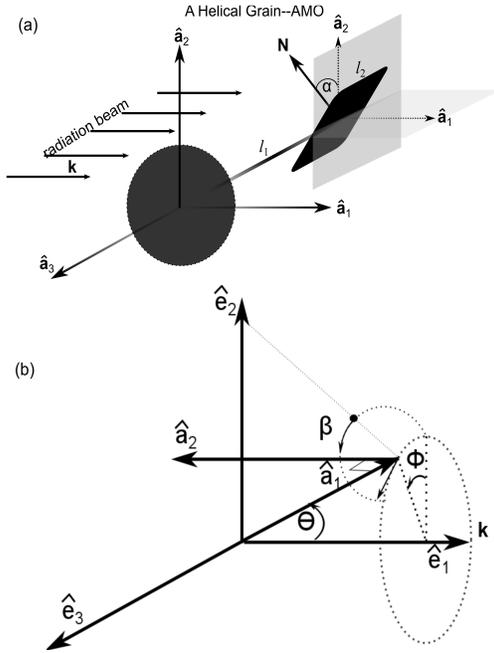

**Figure 1.** (a) A helical grain proposed to analytically compute radiative torques induced by radiation beam **k**: a perfectly reflecting oblate spheroid connected with a weightless mirror by a rod $l_1$. $\mathbf{a}_1$ is the grain's symmetry axis, $\mathbf{a}_2$ and $\mathbf{a}_3$ are two principal axes perpendicular to $\mathbf{a}_1$. $\mathbf{a}_1\mathbf{a}_2\mathbf{a}_3$ defines a system of reference fixed to the grain. $l_2$ is the length and **N** is the normal vector of the mirror. **N** lies in the plane $\mathbf{a}_1\mathbf{a}_2$ and makes an angle $\alpha$ with $\mathbf{a}_2$. (b) Orientation of the grain in the scattering coordinate system $\mathbf{e}_1\mathbf{e}_2\mathbf{e}_3$: $\Theta$ is the angle between $\mathbf{a}_1$ and **k**, $\Phi$ is the precession angle which is measured between $\mathbf{a}_1$ and the plane $\mathbf{e}_1\mathbf{e}_2$, and $\beta$ is the angle between $\mathbf{a}_2$ and $\mathbf{e}_2$. Adapted from LH07a.

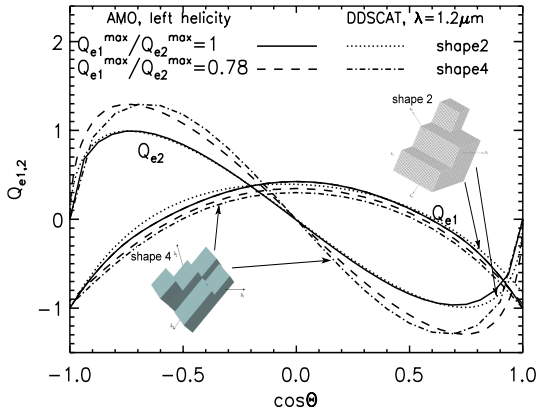

**Figure 2.** Two components of radiative torques, $Q_{e1}$, $Q_{e2}$, as functions of $\cos\Theta$ for $\Phi=0$. Results computed by DDSCAT for two irregular shapes (shape 2 and 4) are shown for comparison (adapted from LH07a).

The functional dependences of torques $Q_{e1}(\Theta)$ and $Q_{e2}(\Theta)$ were shown to be very similar for the analytical model in Fig. 2 and for numerical models of irregular grains subject to radiation of different wavelengths. In Fig.2 this correspondence is shown for two irregular grains (Shape 2 and Shape 4 in LH07a) and AMO.

This remarkable correspondence is also illustrated in Fig. 3a using the function:

$$\Delta^2(Q_{ei}) = \frac{1}{\pi\left(Q_{ei}^{\max}\right)^2} \int_0^\pi \left[Q_{ei}^{irr}(\Theta) - Q_{ei}^{\mathrm{mod}}(\Theta)\right]^2 d\Theta, \quad (1)$$

which characterizes the deviation of the torque component $Q_{ei}$ calculated numerically for irregular grains, from the analytical prediction in the LH07a model.

It is seen in Fig. 3a that while the functional dependence of torque components $Q_{e1}(\Theta)$ and $Q_{e2}(\Theta)$ coincide for grains of various shapes, their amplitudes vary for different grains and different radiation wavelengths (see LH07a). At the same time the ratio of the torque component amplitudes does not change. In fact, LH07a showed that radiative torque alignment can be fully determined if the ratio $q^{\max} = Q_{e1}^{\max}/Q_{e2}^{\max}$ is known. This enormously simplifies the calculations of radiative torques: rather than calculating two *functions* $Q_{e1}(\Theta)$ and $Q_{e2}(\Theta)$ it is enough to calculate two *values* $Q_{e1}^{\max}$ and $Q_{e2}^{\max}$. According to LH07a the maximal value of the function $Q_{e1}(\Theta)$ is achieved for $\Theta=0$, and for the function $Q_{e2}(\Theta)$ it is achieved at $\Theta=\pi/4$. In other words, one can use a *single number* $q^{\max} = Q_{e1}^{\max}/Q_{e2}^{\max} = Q_{e1}(0)/Q_{e2}(\pi/4)$ instead of *two functions* to characterize grain alignment. Thus, it is possible to claim that the $q^{\max}$ ratio is as important for the alignment as the grain axis ratio for producing polarized radiation by aligned grains.

Studying RAT alignment, LH07a corrected the treatment of grain dynamics in Draine and Weingartner (1997). That paper had not treated crossovers, which are essential moments of grain evolution (Spitzer & McGlynn 1979, Lazarian & Draine 1997). A crossover occurs when the grain spin slows down, the grain flips, and thereafter is re-spun up (see Sec. 6.4). When the crossovers are taken into account, the dynamics of the grains becomes very component cannot produce the RAT alignment, as the helicity of an ellipsoidal grain is zero.

different. For instance, instead of accelerating grain rotation, RATs were slowing them down. This slowing down, in realistic circumstances, e.g. in the presence of grain wobbling induced by thermal fluctuations (see Lazarian & Roberge 1997, Hoang & Lazarian 2008), does not bring grains to a complete stop, but results in creating a low-J attractor point. This analytical finding of LH07a was in agreement with an earlier empirical study in Weingartner & Draine (2003).

As a result, apart from high-J attractor points, LH07a found that, for a range of $q^{max}$ and $\psi$ - the angle between the radiation beam and magnetic field direction, only low-J attractor points exist. A later study HL08 established that when low-J and high-J attractor points co-exist, the high-J points are more stable and therefore external stochastic driving, e.g. arising from gaseous bombardment, eventually brings the grains to the high-J attractor points. This transfer can take several damping times, but when a high-J attractor point exists, one can safely assume that the grains are aligned with high angular momentum.

When does alignment happen at low-J attractor points? Fig. 3b shows predictions for the existence of low-J and high-J attractor points by the analytical model (AMO) for the parameter space given by $q^{max}$ and $\psi$. Individual horizontal lines correspond to particular grain shapes with specific values of $q^{max}$. For the interstellar radiation field (ISRF) the calculations of $q^{max}$ were performed using torques averaged over the entire spectrum of radiation field (see LH07a for details). We see that the correspondence in terms of predicting the distribution of high-J and low-J attractor points is also good, which, however, is not surprising due to the good correspondence between the functional dependences obtained for the AMO and irregular grains depicted in Fig. 2.

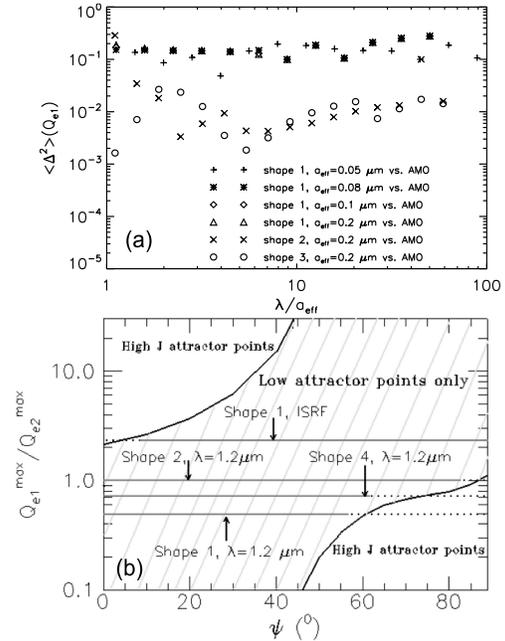

**Figure 3.** (a) Comparison of the torques calculated with DDSCAT for irregular grains for different wavelengths and the AMO of a helical grain. The quantity $\Delta^2$ is defined by Eq. 1. (b) Parameter space for which grains can be aligned with only low-J attractor point (area marked with diagonal lines) and with both low-J and high attractor point (white areas) predicted by AMO. Exact results for irregular grains (shape 1, 2, and 4) obtained using DDSCAT are also shown. When the high-J attractor point is present, grains eventually get there and demonstrate perfect alignment. When only the low-J attractor point is present, the alignment is partial. (From LH07a).

The torques $Q_{e1}$ and $Q_{e2}$ vary with the wavelength of the illuminating field. This results in a change of $q^{max}$ (see Fig. 4a), and the magnitude of the radiative torques. For the latter, LH07a presented a simple numerical fit shown in Fig. 4b. This fit together with the analytical description of RATs in AMO, substantially simplifies modeling of the RAT alignment.

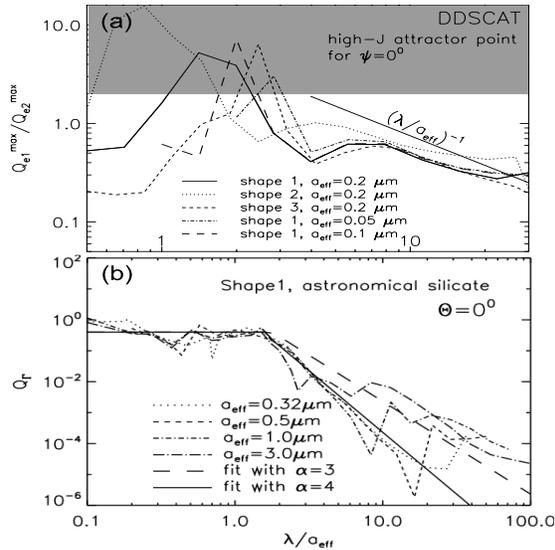

**Figure 4.** (a) The magnitude of the ratio $q^{max} = Q_{e1}^{max}/Q_{e2}^{max}$ that characterize the radiative torque alignment of grains depends on both grain shape and the wavelength of radiation. (b) Torque efficiency for grains of different sizes and wavelengths. The torque amplitude is proportional to radiation intensity. The most efficient alignment is for grains larger than $\lambda/2$. The alignment of grains substantially smaller than the radiation wavelength can also be present if the radiation is strong enough. (From LH07a).

## 3 Properties of RAT alignment

Observations indicate that interstellar grains tend to get aligned with their long axes perpendicular to the ambient magnetic field; a fact that has been used to support the Davis-Greenstein (1951) mechanism responsible for the alignment. Can radiative torques also reproduce this observational fact?

Fig. 5 illustrates why radiative torques tend to align the grains the 'right way', i.e. in agreement with observations. Interstellar grains experience internal relaxation that tends to make them rotate about their axis of maximal moment of inertia. Therefore, it is sufficient to follow the dynamics of angular momentum to determine grain axes alignment. Let us call the component of torque parallel to **J** the *spin-up torque* **H**, and the component perpendicular to **J** the *alignment torque* **F**. The angular momentum **J** is precessing about the magnetic field **B** due to the magnetic moment of the grain, introduced by Barnett magnetization (see Dolginov and Mytrophanov 1976). The alignment torques **F** are perpendicular to **J** and therefore, as **J** gets closer to parallel to **B** (i.e., the angle $\xi$ between **J** and **B** goes to zero), the fast precession of the grain makes these torques averaged to zero. Thus, the positions corresponding to **J** aligned with **B** are stationary points, irrespectively of the functional forms of radiative torques, i.e. of components $Q_{e1}(\Theta)$ and $Q_{e2}(\Theta)$. In other words, the grain can stay aligned with $\xi=0$ or $\xi=\pi$.

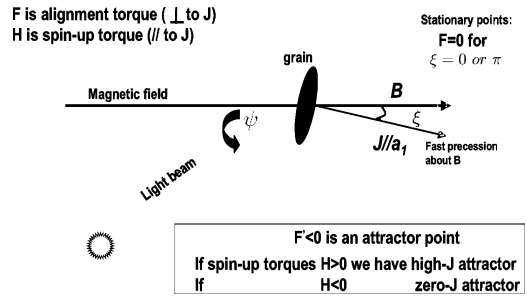

**Figure 5.** A simplified explanation of the grain alignment by radiative torques. The grain, which is depicted as a spheroid in the figure, in fact, should be irregular to get non-zero radiative torque. As shown in LH07a, the positions **J** parallel (or anti-parallel) to **B** correspond to the stationary points as at these positions the component of torques that changes the alignment angle vanishes, and the grain gets aligned with long axis perpendicular to **B**.

The arguments above are quite general, but they do not address the question whether there are other stationary points, e.g. whether the alignment can also happen with **J** perpendicular to **B**. To answer this question one should use the actual expressions for $Q_{e1}(\Theta)$ and $Q_{e2}(\Theta)$. The analysis in LH07a shows that there is, indeed, a range of angles between the directions of the radiation and the magnetic field for which grains tend to align in a 'wrong' way, i.e. with their long axes parallel to the magnetic field. However, this range of angles is rather narrow and does not exceed several degrees. Grain thermal wobbling at low-J attractor point (Lazarian 1994, Lazarian and Roberge 1997) induces variations in the angle that typically exceeds this range.

**Table 1: Timescales relevant for grain alignment**

| Symbol | Meaning | Definition | Value (s) |
|---|---|---|---|
| $t_{rot}$ | Thermal rotation period | $2\pi/\omega_T$ | $2.7\times 10^{-5}\hat{\rho}^{1/2}\hat{T}_g^{-1/2}\hat{s}^{1/2}a_{-5}^{5/2}$ |
| $t_{gas}$ | Gas damping time | $\dfrac{3I_\parallel}{4\sqrt{\pi}nm_H v_{th}a^4}$ | $1.6\times 10^{12}\hat{\rho}\hat{n}^{-1}\hat{T}_g^{-1/2}\hat{s}a_{-5}$ |
| $t_{Bar}$ | Barnett relaxation time | $\dfrac{\gamma_e^2 I_\parallel^2}{VK_e(\omega_1)h^2(h-1)J^2}$ | $1.5\times 10^{7}\left(\dfrac{\hat{\rho}^2}{\hat{T}_d\hat{K}_e}\right)f_1(s)a_{-5}^7\left(\dfrac{J_d}{J}\right)^2 F(\omega_1,\tau_e)$ |
| $t_{nucl}$ | Nuclear relaxation time | $\left(\dfrac{\gamma_n}{\gamma_e}\right)^2\left(\dfrac{K_e}{K_n}\right)t_{Bar}$ | $7.6\times 10^{1}\left(\dfrac{\hat{\rho}^2}{\hat{T}_d\hat{K}_n}\right)\hat{g}_n^4\hat{\mu}_n^{-2}f_1(s)a_{-5}^7\left(\dfrac{J_d}{J}\right)^2 F(\omega_1,\tau_n)$ |
| $t_{tf}$ | Thermal flipping time |  | $t_{Bar,nucl}\exp\left(0.5\left[(J/J_d)^2-1\right]\right)$ |
| $t_{DG}$ | Paramagnetic alignment time | $\dfrac{I_\parallel}{VK_e(\omega)B^2}$ | $4.0\times 10^{13}\hat{\rho}\hat{B}^{-2}\hat{T}_d\hat{K}_e^{-1}a_{-5}^2 F(\omega,\tau_e)$ |
| $t_c$ | Crossover time | $\dfrac{2J_d}{L_z^b}$ | $1.6\times 10^{9}\left(\dfrac{\hat{\rho}\hat{T}_d\hat{\alpha}}{\hat{W}\hat{\zeta}\hat{n}^2\hat{T}_g}\right)^{1/2}f_2(s)a_{-5}^{1/2}$ |
| $t_L$ | Larmor precession time | $\dfrac{2\pi I_\parallel\omega}{\mu_{Bar}B}$ | $2.7\times 10^{7}\left(\dfrac{\hat{\rho}}{\hat{\chi}\hat{B}}\right)a_{-5}^2$ |
| $t_{RAT}$ | Radiative precession time | $\dfrac{2\pi}{\lvert d\phi/d\lambda\rvert}$ | $\dfrac{7.3\times 10^{10}}{\hat{Q}_{e3}}\hat{\rho}^{1/2}\hat{T}_g^{1/2}\hat{s}^{1/6}a_{-5}^{1/2}\left(\dfrac{1}{\hat{\lambda}\hat{u}_{rad}}\right)$ |
| $t_E$ | Electric precession time | $\dfrac{2\pi J}{p.E}$ | $1.8\times 10^{4}\hat{\varepsilon}\hat{U}\hat{E}^{-1}\hat{\rho}^{1/2}\hat{\omega}\hat{T}_g^{1/2}a_{-5}^{1/2}$ |

| Notations | | | |
|---|---|---|---|
| a, b: lengths of semimajor and semiminor axes of the oblate spheroidal grain | $s=b/a$: axial ratio, $a_{-5}=a/10^{-5}$cm, $\hat{s}=s/0.5$; $\rho=3\text{gcm}^{-3}\hat{\rho}$: mass density of grain material | $h=\dfrac{I_\parallel}{I_\perp}$: ratio of inertia moments $\omega$: grain angular velocity | $\omega_T=(k_B T/I_\parallel)^{1/2}$: thermal angular velocity |
| $T_g=100\text{K }\hat{T}_g$: gas temperature | $T_d=15\text{K }\hat{T}_d$: dust temperature | $T_{rot}=1/2(T_d+T_g)$: rotational temperature | $n=30\text{cm}^{-3}\hat{n}$: gas density |
| $B=5\mu\text{G }\hat{B}$: magnetic field | $\chi(0)=\hat{\chi}/10^{-4}$: magnetic susceptibility at zero rotation frequency | $\omega K_{e,n}$: imaginary part of susceptibility for electrons and nuclei | $K_e=1.2\times 10^{-13}\text{s }\hat{K}_e$ $K_n=10^{-15}\text{s }\hat{K}_n$ |
| $\gamma_e=\dfrac{g_e\mu_B}{h/2\pi}$: gyromagnetic ratio for electron | $\gamma_n=\dfrac{g_n\mu_N}{h/2\pi}$: gyromagnetic ratio for nucleus | $\mu_e=\dfrac{eh}{4\pi m_e c}$, $\mu_N=\dfrac{eh}{4\pi m_p c}$: Bohr magneton | $\mu_n=\hat{\mu}_n\mu_N$: nuclear magnetic moment |
| $J_d=[I_\parallel kT_d/(h-1)]^{1/2}$: $\perp$ thermal angular momentum at $T=T_d$ | $t_{Bar,nucl}^{-1}=t_{Bar}^{-1}+t_{nucl}^{-1}$: total internal relaxation rate | $u_{rad}=\hat{u}_{rad}u_{ISRF}$: energy density of radiation field; $u_{ISRF}$: energy density of local interstellar radiation field | $\lambda=1.2\mu\text{m}\hat{\lambda}$: wavelength of radiation field |
| $Q_{e3}=10^{-2}\hat{Q}_{e3}$: third component of radiative torques | $E=10^{-5}\text{Vcm}^{-1}\hat{E}$: electric field $\hat{\omega}=\omega/\Omega_T$: normalized rotation velocity | $p=10^{-15}\hat{U}\hat{\varepsilon}a_{-5}$ statC cm: electric dipole moment; $U=0.3\text{V }\hat{U}$: grain potential | $\varepsilon=10^{-2}\hat{\varepsilon}$: displacement of grain charge centroid from its center of mass |
| $L_z^b$: magnitude of $H_2$ torque | $\zeta=0.2\hat{\zeta}$: $H_2$ formation efficiency | $W=0.2\text{eV }\hat{W}$: kinetic energy of ejected $H_2$ | $\alpha=10^{11}\text{cm}^{-2}\hat{\alpha}$: density of recombination site |
| $F(\omega_1,\tau)=[1+(\omega_1\tau/2)^2]^{-2}$ $\omega_1=(h-1)J\cos\theta/I_\parallel$ | $\tau_e,\tau_n$: electronic and nuclear relaxation times; $\theta$: angle between J and grain symmetry axis | $f_1(s)=\hat{s}\left(\dfrac{1+s^2}{1.25}\right)^2$ | $f_2(s)=\left(\dfrac{1+s^2}{s(1-s^2)}\right)^{1/2}$ |

Thus a remarkable result emerges: in RAT alignment, grains always get aligned with their long axes perpendicular to **B**! The fact that observations indicate that grains must be aligned with their long axes perpendicular to the magnetic field was at one time used to "prove" that the alignment is paramagnetic. We now see that RATs produce the same sense of alignment. However, the strength for typical interstellar magnetic fields is much larger than that for paramagnetic torques.

One of the important features of RAT alignment predicted by AMO is that the phase trajectory map is symmetric with respect to grain flipping state (i.e., initial direction between the axis of maximal inertia $a_1$ and grain angular momentum **J**), and that the positions of low-J and high-J attractor points in the trajectory map do not depend on the initial flipping state (see Fig. 6b). In other word, flipping the grain by 180° in the fixed ambient radiation does not significantly change the dynamics of the grain (i.e., grain alignment). This feature, which appears to be counterintuitive, results from the grain helicity, and that helicity is invariant on reflection.

Consider an arbitrary point P in the phase trajectory of grains with the positive flipping state, i.e., $a_1$ parallel to **J** (squares in Fig..6b). This grain state corresponds to $\cos\Theta=-1$ or $\Theta=180°$ ($\Theta$ is the angle between $a_1$ and **k** or **B**). The grain is being decelerated by RATs so that it will be driven to the attractor point C. Now, the grain is instantaneously flipped by 180 degree (e.g., rotating the grain around the $a_2$ axis) such as $a_1$ becomes antiparallel to the fixed **J**, then, this new state of the grain is shown by the point N in Fig. 6(b), with $\cos\Theta=-1$ and $\Theta=0$. At the point N, one can see that the grain is also being decelerated by RATs and eventually is driven to the attractor point C. Thus, the attractor point C is shared by grains in both flipping states.

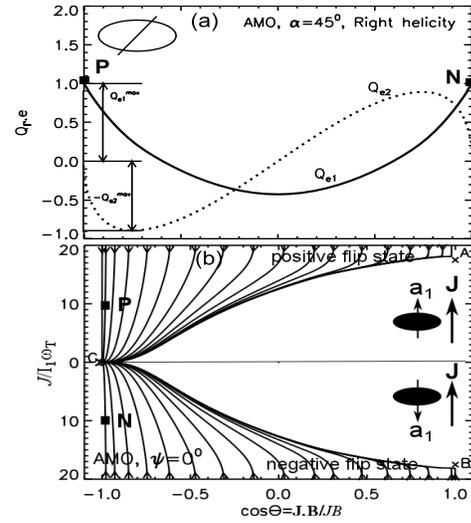

**Figure 6.** (a) Two torque components $Q_{e1}$ and $Q_{e2}$ from AMO. (b) A typical trajectory phase map describing the evolution of grain momentum, $J/I_1\omega_T$, and cosine of angle between **J** and the ambient magnetic field subject to RATs and rotational gas damping from AMO for the case the external magnetic field **B** parallel to the anisotropic radiation direction. Initially, grains are assumed to be in positive flipping state ($a_1$ parallel to **J**) and negative flipping state ($a_1$ antiparallel to **J**) and then they are driven by RATs to low-J attractor points (C) and high-J repeller points (A and B). Results from HL08.

To see why this happens, let us look closely at the RATs in Fig. 6a. It can be seen that when the grain flips from the points P ($\Theta=180°$) to N ($\Theta=0°$), $Q_{e1}(\Theta)$ does not change due to its symmetry and $Q_{e2}(\Theta)=0$ at $\Theta=0$ and $180°$ (Fig. 6a). As a result, the spin-up torque component $H=\cos\Theta\ Q_{e1}(\Theta)$ remains the same under the grain flip, and the grain dynamics does not change.

Physically, considering the helical grain with initial orientation as in Fig. 1a, one can see that the front surface of the mirror (surface 1) is directly illuminated by the radiation beam while the back surface (surface 2) is not, and the grain is accelerated along the radiation direction **k**. As the grain flips (rotation by 180° along $e_3$ parallel to $a_3$ axis), it is easily seen that the angle between the mirror normal vector **N** and **k** is the same, and the surface 2 of the mirror becomes directly illuminated by the radiation beam. Since two mirror surfaces are assumed to be identical, RATs arising from the reflection by the two surfaces are the same for the same angle between the mirror plane and the radiation direction. As a result, the grain is still accelerated in the radiation direction **k** after the flipping. If the two mirror surfaces have different reflection coefficients, the resulting RATs would become different for two flipping states. The fact that RATs do not change with the grain flipping is due to its helical nature, which demonstrates the RAT is always directed along the anisotropic direction of radiation.

**4 Complexities of dust dynamics**

Grain dynamics is rather complex and it must be accounted for in quantitative modeling, aimed at direct comparisons with observations.

*4.1 Effects of internal relaxation*

Grain alignment is a complex process. To produce the observed starlight polarization, grains must be aligned, with their *long axes* perpendicular to the magnetic field. The degree of alignment involves alignment not only of the grain's angular momentum **J** with respect to the external magnetic field **B**, but also the alignment of the grain's short axis with respect to **J**. The latter alignment is frequently termed the "internal alignment".

Purcell (1979) was the first to consider internal relaxation as the cause of internal alignment. He showed that the rates of internal relaxation of energy in wobbling grains are much faster than the time scale for grain alignment. This induced many researchers to think that all grains rotate about the axis of their maximal inertia moment and the internal alignment is perfect. Lazarian (1994) corrected this conclusion by showing that the fluctuations associated with the internal dissipation, required by the Fluctuation Dissipation Theorem, should induce wobbling, the amplitude of which increases as the effective rotational temperature of grains decreases and approaches that of the grain material (see also Lazarian and Roberge 1997).

The rate of internal relaxation is an important characteristic of grain dynamics. Spitzer and McGlynn (1979), Lazarian and Draine (1997, 1999a,b), LH08, HL09a demonstrated its crucial significance for various aspects of grain alignment. Purcell (1979) was the first to evaluate the rates of internal relaxation, taking into account the inelastic effects and a particular effect that he discovered himself, the Barnett relaxation. The Barnett effect is the magnetization of a paramagnetic body as the result of its rotation. Purcell (1979) noted that a wobbling rotating body would experience changes of the magnetization arising from the changes of the direction of rotation in grain axes. This would entail fast relaxation, with a characteristic time of about a year for $10^{-5}$ cm grains. Lazarian and Draine (1999a) identified $10^6$ times stronger relaxation related to nuclear spins and LH08 showed that superparamagnetic grains, e.g. grains with magnetic inclusions (see Jones and Spitzer 1967, Mathis 1986, Martin 1995), also exhibit the enhanced internal relaxation.

The relative role of internal relaxation for Barnett, nuclear, and inelastic relaxation is shown in Fig. 6(a). The calculations of inelastic relaxation from Lazarian and Efroimsky (1999) are used for the plot. These rates are important for grain dynamics. In particular, they influence how grain crossovers happen (see Sec. 6.3).

*4.2 Purcell's pinwheel torques*

It is important to understand whether additional torques can influence RAT alignment. Purcell (1975, 1979) realized that grains may rotate at a rate much faster that the thermal rate if they are subject to systematic torques. Purcell (1979) identified three such systematic torque mechanisms: inelastic scattering of impinging atoms when gas and grain temperatures differ, photoelectric emission of electrons, and $H_2$ formation on grain surfaces (Fig. 7b). Below we shall refer to the last of these as "Purcell's torques". These were shown to dominate the other two for typical conditions in the diffuse ISM (Purcell 1979). The existence of systematic $H_2$ torques is expected due to the random distribution over the grain surface of catalytic sites of $H_2$ formation, since each active site acts as a minute thruster emitting newly-formed $H_2$ molecules. A later study of uncompensated torques in HL09a added additional systematic torques to the list, namely, torques arising from plasma-grain interactions and torques arising from the emission of radiation by an irregular grain. Radiative torques arising from the interaction of the *isotropic* radiation with an irregular grain (Draine and Weingartner 1996) also represent systematic torques fixed in the grain body. Emission of an irregular grain also induces systematic torques (HL09a). We shall call all the above systematic torques *pinwheel torques* to distinguish them from the systematic torques arising from an anisotropic flow of photons or atoms interacting with a helical grain (see below).

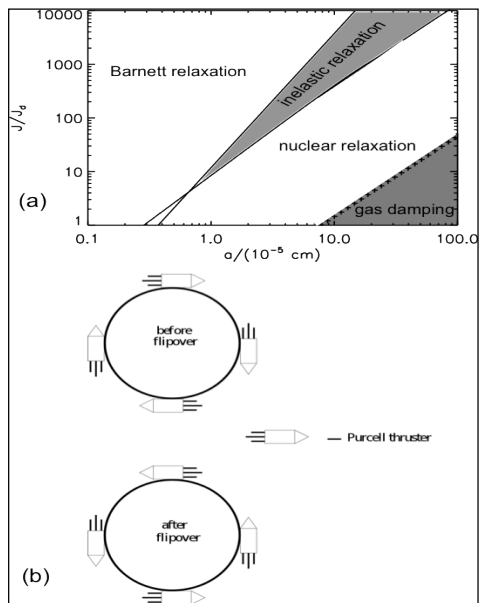

**Figure 7.** (a) The relative role of the Barnett, nuclear and inelastic relaxation for grains of different sizes. Rates of Barnett relaxation are from Purcell (1979), nuclear relaxation are from Lazarian and Draine (1999b) and for inelastic relaxation are from Lazarian and Efroimsky (1999). (b) A grain acted upon by Purcell's torques before and after a flip over event. As the grain flips, the direction of torques alters. The $H_2$ formation sites act as thrusters (from Lazarian (2007); reproduced by permission of Elsevier).

Purcell (1979) considered changes of the grain surface properties and noted that those should stochastically change the direction (in body-coordinates) of the systematic torques. If this happens, the grains get decelerated and undergo eventually a process that Spitzer and McGlynn (1979) termed *crossovers*. During a crossover, the grain slows down, flips, and after the flip is accelerated again (see Fig. 8a).

From the viewpoint of grain-alignment theory, the important question is whether or not grain's angular momentum gets randomized during a crossover. If the value of the angular momentum is small during the crossover, the grains are susceptible to randomization arising from atomic bombardment.

The original calculations in Spitzer and McGlynn (1979) obtained only marginal correlation between the values of the angular momentum before and after a crossover, but this was due to the fact that their analysis disregarded thermal fluctuations within the grain with temperature $T_d$. According to the Fluctuation-Dissipation Theorem, these thermal fluctuations induce thermal wobbling with a frequency determined by the internal relaxation rate (Lazarian 1994). If the crossover happens over a time larger than the grain internal relaxation time, the Spitzer and McGlynn (1979) theory of crossovers should be modified to include the value of thermal angular momentum $J_{d\perp} = (2I_\perp k T_d)^{1/2}$, where $I_\perp$ is the moment of inertia of an oblate grain, and $T_d$ is the dust temperature (Lazarian and Draine 1997).

*4.3 Thermal trapping of grains*

The arguments above are applicable to grains larger than some critical size $a_c$ determined by the grain relaxation efficiency. What happens for grains smaller than $a_c$?

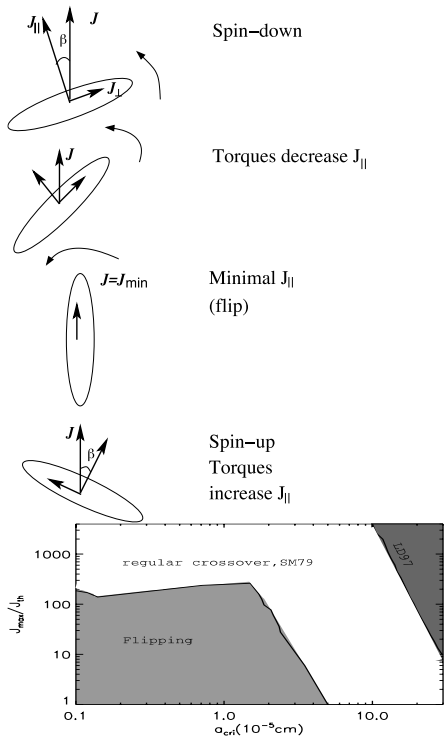

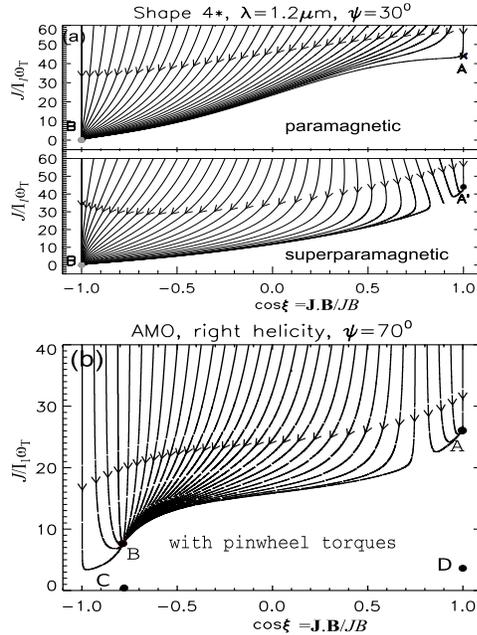

**Figure 8.** (a) Upper panel: a regular crossover event as described by Spitzer and McGlynn (1979). The systematic torques nullify the amplitude of the **J** component parallel to the axis of maximal inertia, while preserving the other component, $J_{\perp}$. If $J_{\perp}$ is small then the grain is susceptible to randomization during crossovers. The direction of **J** is preserved in the absence of random bombardment. Reproduced by permission of Elsevier. (b) Lower panel: the applicability range for the Spitzer and McGlynn (1979), Lazarian and Draine (1997) and Lazarian and Draine (1999a,b) models. The latter corresponds to the shaded area of "Flipping". The corresponding flipping rates were elaborated in HL09a. $J_{max}$ parameterizes here the value of the pinwheel torques, i.e. $J_{max}=t_{gas}dJ/dt$, where the derivative $dJ/dt$ arises from the pinwheel torque action.

Lazarian and Draine (1999a) found that particles smaller than this size flip fast. This new effect will alter the action of Purcell torques, making the grain rotate thermally in spite of the presence of these uncompensated torques, e.g. torques arising from $H_2$ formation (see Fig. 7(b)). Lazarian and Draine termed this phenomenon, suppressing the effect of uncompensated torques, "thermal trapping".

Based on these arguments, Lazarian and Draine (1999a) predicted that interstellar grains smaller than $10^{-6}$ cm must be poorly aligned via paramagnetic relaxation. A more elaborate study of the flipping phenomenon in Roberge and Ford (2000) supported this conclusion.

Weingartner (2009), however, concluded that the spontaneous thermal flipping does not happen if the internal relaxation diffusion coefficient is obtained with the correct integration constant (Lazarian and Roberge 1997, Roberge and Ford 2000). Does this invalidate the phenomena of *thermal flipping* and *thermal trapping* predicted in Lazarian and Draine (1999a)?

**Figure 9.** Phase trajectories of grains in terms of the angular momentum and the cosine of angle between **J** and **B** (from Lazarian and Hoang (2008) and Lazarian (2009a); reproduced by permission of the AAS). (a) A paramagnetic grain gets only a low-J attractor point. For the same set of parameters a super-paramagnetic grain gets also a high-J attractor point. The fact that most of the phase trajectories go in the direction of the low-J attractor point illustrates the dominance of the radiative torques for the alignment even in the case of the super-paramagnetic grain. However, high-J attractor points are more stable than the low-J attractor points. As a result, all grains eventually end up at the high-J attractor point. (b) Grain alignment by radiative torques in the presence of pinwheel torques. The shown case corresponds to the presence of both the low-J and high-J attractor points in the absence of pinwheel torques. In the case when only a low-J attractor point exists the strong pinwheel torques lift the low-J attractor point enhancing the alignment.

The study in Hoang and Lazarian (2009a) shows that the flipping and trapping does happen, if one takes into account additional fluctuations associated, for example, with the action of the uncompensated torques or gaseous bombardment. However, these arguments modify the value of the critical size at which thermal trapping occurs. Therefore, there could exist non-flipping grains with a size smaller than $a_c$. For those grains, the Lazarian and Draine (1997) theory of crossovers is applicable. This theory predicts enhanced paramagnetic alignment. However, such grains will be only marginally aligned via the Davis-Greenstein process if the pinwheel torques are short-lived, i.e. the time scale of their existence is much smaller than the time of paramagnetic relaxation. Such grains may rotate at high rate in accordance with the Purcell (1979) model. As we discuss below, the fact that pinwheel torques are not suppressed by flipping may allow better alignment by RATs.

Fig. 8b defines the range over which the different models of crossovers are applicable. It is clear that for a sufficiently large $J_{max}$ the flipping gets suppressed. The picture above, however, is different if the grains paramagnetically dissipate the energy, via the Davis-Greenstein process, on time scales shorter than the gaseous damping time. LH08 found that if grains have super-paramagnetic inclusions, they *always* get high-J attractor point.

Fig. 9a shows that for super-paramagnetic grains subject to a diffuse interstellar radiation field, most grains still go to the low attractor point, which reflects the fact that radiative torques rather than paramagnetic ones dominate the alignment. As the high-J attractor point is more stable compared to the low-J attractor point, similar to the dynamics for ordinary paramagnetic grains, super-paramagnetic grains get transferred by gaseous collisions from the low-J to high-J attractor points. Thus, super-paramagnetic grains always rotate at a high rate in the presence of radiative torques. One concludes that, rather unexpectedly, intensive paramagnetic relaxation changes the rotational state of the grains, enabling them to rotate *rapidly*. The alignment of super-paramagnetic grains at high-J point is *perfect*, i.e. grains rotate with **J** along **B**. Fig. 9b shows that pinwheel torques increase the alignment by increasing the angular momentum of grains at attractor points.

## 5 Alignment of large grains

The internal relaxation of energy in rotating grains considerably simplifies their dynamics. However, our discussion in Section 6.4 indicates that, for sufficiently large grains, e.g. for grains of the size larger than ~$10^{-4}$cm, the internal relaxation over the time-scale of RAT alignment may be negligible. Large grains are present e.g. in accretion disks and comets. The polarization from accretion disks can provide one with an important insight into the magnetic fields of the objects (see Cho and Lazarian 2007), while understanding of the alignment for comet dust is important for predicting circular polarization (see below). Thus, a proper description of RAT alignment in the absence of internal relaxation is important.

In the absence of internal relaxation, grains can get aligned not only with their long axes perpendicular to the angular momentum **J**, but also with the long axis parallel to **J**, (i.e. the axis of minimal moment of inertia parallel to **J**). This complicates the analysis compared to the case of interstellar grains, for which the internal relaxation is overwhelming. The corresponding problem was addressed in HL09b. The results of this study are summarized in Table 2.

**Table 2.** Attractor points with and without internal relaxation for AMO. "L" denotes "long axis of the grain."

| Without relaxation (HL09b) | | With relaxation (LH07a) | |
|---|---|---|---|
| High-J | Low-J | High-J | Low-J |
| J∥B | J∥ or at angle with B | J∥B | J∥ or at angle with B |
| L⊥B | L⊥ or ∥J | L⊥B | L⊥B |

Hoang and Lazarian (2009b) suggest that grains do preferentially get aligned with their long axes perpendicular to the magnetic field even without internal relaxation. Indeed, this is consistent with the finding that such alignment happens for the high-J attractor points of AMO. If this is the case, when the low-J and high-J attractor points coexist, the steady-state alignment will happen only at high-J attractor points (see earlier discussion). The wrong alignment happens only at low-J attractor points and it is suggestive that in the presence of gaseous bombardment the grains may still spend more time in the vicinity of the high-J repeller point, as was shown in HL08. The conclusions obtained with AMO are consistent with HL09b for one of the irregular grains. However, more extensive studies of the RAT alignment in the absence of internal relaxation are required.

**6 Astrophysically important implications**

*6.1 Molecular clouds and diffuse/dense cloud interface*

Aligned grains provide a very cost-effective way to trace magnetic fields in molecular clouds and getting insight into the role of magnetic field in star formation (see Hildebrand et al. 2000, Crutcher 2012). In dense molecular clouds one might expect that all grain alignment mechanisms are suppressed as conditions there approach those of thermal equilibrium (see discussion in Lazarian et al. 1997). This conclusion, however, contradicts observations in Ward-Thompson et al. (2000), which revealed aligned grains in dense pre-stellar cores with total visual extinction up to $A_V$ =15. Cho and Lazarian (2005) noticed that the efficiency of grain alignment is a function of the grain size (see Fig. 4b), and this allows even highly attenuated reddened interstellar light to align the larger grains that are known to exist in dark cores. More detailed calculations (Bethell et al. 2007) supported this explanation. However, even in the latter paper, which considered the irregularity of density distribution for the purposes of radiation transfer, only crude estimates of the grain alignment efficiencies were used. Future studies should capitalize on the recent progress of grain alignment theory initiated by the LH07a study. For instance, the attenuated light in molecular clouds can be decomposed into the dipole and quadrupole components and it is advisable to take into account the theoretical predictions for the alignment by these components obtained in Hoang and Lazarian (2009b). Comparing Figs, 3b and 10a one can see that the alignment by the dipole component corresponds to a larger range of angles with high attractor points, which also means perfect alignment for more grains.

Modeling of the variations of the dependence of the polarization degree with wavelengths - referred to by Hildebrand et al. (2000) as the "polarization spectrum" - is another important test of grain alignment. Calculations of the polarization spectrum in the absence of embedded stars are presented in Lazarian (2007), while those in the presence of stars are made in our recent study. The latter exhibit better correspondence with most observations, which is not surprising as the clouds with embedded stars were studied in Hildebrand et al. (2000).

In terms of polarization in diffuse media, the RAT alignment reproduces the empirical Serkowski (1973) law well (see Lazarian 2007 and section 6.8.1). Moreover, RATs can explain the change of the polarization degree observed at the interface of the diffuse media and a molecular cloud. The latter variations were first reported in Whittet et al. (2001) and explained as a consequence of the RAT-induced alignment in Lazarian (2003). A more extended data set was analyzed in Whittet et al. (2008), where a quantitative correspondence between the RAT predictions and the observations was established. A fit to Whittet et al. (2008) data, which, apart from the RATs efficiencies, takes into account the effects of magnetic field turbulence is shown in Fig. 10b.

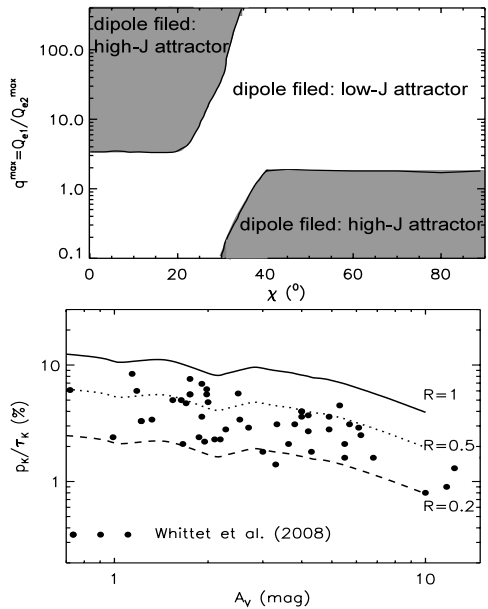

**Figure 10.** (a) Upper panel: tthe parameter space similar to that in Fig. 2b, but for the external dipole radiation (from HL09b). (b) Lower panel: theoretical predictions for the RAT alignment in the presence of turbulence compared with Whittet et al. (2008) data. R is the Rayleigh reduction factor, which characterizes the alignment. Evidently, the alignment of 20 % is too weak to explain the data. Both figures are reproduced by the permission of AAS.

### 6.2 Polarization of Zodiacal light

Circular polarization of Zodiacal light was reported by Wolstencroft and Kemp (1972). Those authors explained this by scattering of unpolarized solar radiation by aligned grains. The mechanism of alignment was not addressed. Later Dolginov and Mytrophanov (1976) mentioned possible alignment of Zodiacal dust. Nevertheless, the alignment of Zodiacal dust has been mostly ignored and now presents a possible problem for studies of polarized Cosmic Microwave Background (CMB) emission.

In Hoang and Lazarian (2014) we present theoretical calculations of the expected circular polarization $P_V/P_V^{max}$ as a function of elongation angle $\varepsilon$ between the direction from the Earth to the Sun and the line of sight, for an intentionally simplified model of the interplanetary magnetic field. The direction of magnetic field is assumed to incline by various angles with the ecliptic plane, and the component parallel to the ecliptic plane makes a constant angle with respect to the all lines of sight. The circular polarization $P_V/P_V^{max}$ appears to change its sign at $\varepsilon$ =90, 180, 220, 270 and 320 degree (see Chapter IV. 24 for details). By comparing $P_V/P_V^{max}$ from calculations with observation data, one can infer the magnetic field geometry in the interplanetary medium.

Could there be no circular polarization of Zodiacal light and a substantial polarized contribution from the Zodiacal dust to the CMB emission? The answer to this question is yes. Indeed, grain alignment theory predicts efficient grain alignment in the solar neighborhood. If magnetic fields are exactly in the plane of ecliptic, then the circular polarization is zero. Indeed, the circular polarization is proportional to the scalar product of the magnetic field direction $\mathbf{B}/|\mathbf{B}|$ and the vector product of $\mathbf{n}_E \times \mathbf{n}_S$, where $\mathbf{n}_E$ gives the direction from the scattering volume to the Earth and $\mathbf{n}_S$ provides the direction from the Sun to the scattering volume. When all three vectors are in the same plane, the result is zero.

The measurements in Wolstencroft and Kemp (1972) show that in most cases the magnetic field is oriented at an angle to the ecliptic, which provides a means of studying the variations of the magnetic field both in space and time. The long-range variations of the magnetic field should provide input on the injection scale of Solar wind turbulence, which is by itself an important problem.

### 6.3 Alignment of dust in comets

Polarization arising from comets is described in detail in Chapter IV.22. We would like only to stress that the changes in the polarization of stars during their occultation by comets C/1990 K1 (Levy) and C/1995 O1 (Hale-Bopp) that are described in the chapter clearly show the presence of aligned grains in comet atmospheres. This aligned dust can produce the circular polarization that has been also observed (Rosenbush et al. 2007). The latter can be explained by RATs, provided that the axis of alignment is not given by the radiation flow. Magnetic fields can provide such an axis at sufficient distances from the comet nucleus, but it is generally believed that there are no magnetic fields close to the nucleus. In this situation, electric fields of comets (Serezhkin 2000) can act as the alignment axes if the grain electric moment is not zero (see Draine and Lazarian 1998b). Another possibility is related to the gaseous flow inducing grain precession. It was discussed in Lazarian and Hoang (2007b) that irregular grains may be aligned by uncompensated mechanical torques similar to RATs. Our more recent study shows that such mechanical torques are less efficient compared to RATs due to averaging of grain helicity as different facets of grain are being subject to gaseous bombardment. Nevertheless the $Q_3$ component of the torques that induces grain precession does not depend on grain helicity, enabling the flow to be the alignment axis. This entails the prediction of RAT aligned grains with the axis of alignment far from the comet being given by the magnetic field and with different process defining the axis of alignment for grains near the comet nucleus. Note that paramagnetic alignment cannot possibly align grains close to a comet nucleus.

## 7 Alignment by paramagnetic relaxation

RAT alignment mechanism has been proven to be robust for interstellar grains in various environment conditions, and its predictions are consistent with observational data. However, it appears to be inefficient for small and very small grains because RATs decrease rapidly with the decreasing grain size as $(\lambda/a)^{-3}$ for $a<<\lambda$ (LH07a, see also Fig. 4b). Yet observational data and modeling (see Hoang et al. 2013 and references therein) show that a finite degree of alignment of small interstellar grains is needed for reproducing the observed data. In this Section, we discuss the alignment of small grains by paramagnetic relaxation and its implications.

*7.1 Alignment of small grains*

Kim and Martin (1995) employed maximum entropy method to fit theoretical polarization curves to observational data and discovered that interstellar silicate grains of size a>0.05μm are efficiently aligned, which is consistent with the alignment driven by RATs. Moreover, they found some residual alignment for small *(a ~ 0.01μm)* grains.

Recently, motivated by polarized submillimeter emission measured by the *Planck* mission, Draine and Fraisse (2009) derived the alignment function for interstellar grains by fitting simultaneously the observed extinction and polarization curves for the typical diffuse ISM ($R_V$=3.1, $\lambda_{max}$=0.55μm). They came to the same conclusion as Kim and Martin (1995) that for a > 0.05μm grains are efficiently aligned. In addition, they found that the degree of alignment for small grains is f ~ 0.01, with only silicate grains aligned.

Observational evidence for residual alignment of small grains is numerous. For instance, Clayton et al. (1992, 1995) reported an excess polarization in the ultraviolet (UV) in comparison with the Serkowski law (Serkowski et al. 1975) extrapolation for a number of stars with low $\lambda_{max}$ (e.g., $\lambda_{max}$ <0.55μm). They found a tight correlation between the excess in the UV polarization characterized by P(6μm$^{-1}$)/P$_{max}$ and 1/$\lambda_{max}$. In particular, they showed no difference in the UV extinction for the stars with a UV polarization excess, indicating that the properties of the dust are not unusual. Using updated observational data, Martin et al. (1999) have confirmed the relationship between P(6μm$^{-1}$)/P$_{max}$ and 1/$\lambda_{max}$ and showed that the UV polarization can be described by a Serkowski-like relation. A systematic change in the mass distribution of aligned grains was suggested as a potential cause of the relationship (Martin et al. 1999), although at that time the alignment mechanism, responsible for the effect, was not known.

Could the UV polarization excess arise from enhanced RAT alignment by nearby stars? The answer is inconclusive. To date, the stars with observed UV polarization excess do not have emission excess at 60μm (see Clayton et al. 1995), which indicates that dust along these lines of sight is actually not hotter than dust along the line of sight to the stars without UV polarization excess. Although excess UV from hot white dwarfs or subdwarf stars along the line of sight to the target stars cannot be ruled out as causing the alignment.

Hoang et al. (2014) fitted theoretical models with aligned silicates to observed extinction and polarization curves for the stars with $\lambda_{max}$ <0.55μm. They found that the degree of alignment of small grains tends to increase with decreasing $\lambda_{max}$ (Fig. 11). They attributed the enhanced alignment of small grains to an increase in the magnetic field strength. Combining theoretical calculations for the degree of alignment of small grains by paramagnetic relaxation with best-fit models, they found that the upper limit of the interstellar magnetic field is B~ 10μG for the typical ISM of $\lambda_{max}$=0.55μm and is required to increase to B~ 15μG to reproduce the UV polarization excess for the cases of low $\lambda_{max}$=0.51μm.

If the enhanced alignment of small grains is indeed due to the increase of magnetic field strength, then one important implication of UV polarization is that it allows us to constrain *the strength of the magnetic field*. For instance, Hoang et al. (2014) show that, from the three observational polarization parameters $P_{max}$, $\lambda_{max}$ and P(UV), one can estimate the magnetic field strength. This new technique can be complementary to other existing techniques (e.g., Zeeman effect, Chandrasekhar-Fermi technique, atomic alignment (see Yan and Lazarian 2006) and it can help to obtain a better constraint for the interstellar magnetic field.

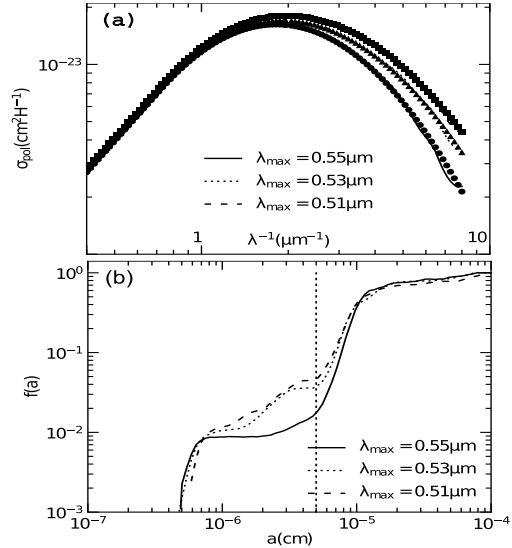

**Figure 11.** (a) Observed polarization curves (● for 0.55mm, ▲ for 0.53 mm, and ■ for 0.51 mm) and best-fit models for three values of $\lambda_{max}$ and $P(\lambda_{max})/A(\lambda_{max})$=3% mag. (b) Alignment functions for our best-fit models. The alignment of small grains (a<0.1μm) increases with the decreasing $\lambda_{max}$. Reproduced from Hoang et al. (2014) by permission of the AAS.

*7.2 Alignment of ultrasmall grains and polarized spinning dust*

The problem of alignment of ultra-small grains (e.g., PAHs) is crucial for understanding the polarization of spinning dust emission (Draine and Lazarian 1998a,b), an important component of Galactic foregrounds that dominates the CMB signal (see Planck Collaboration 2013). The first study exploring the paramagnetic alignment of PAHs was performed by Lazarian and Draine (2000). They identified a new process of relaxation, which was termed "resonance paramagnetic relaxation" and which is orders of magnitude more efficient for the rapidly rotating PAHs compared to the classical Davis-Greenstein process. However, in the last several years, significant progress has been made in understanding spinning dust in terms of both theory (Hoang et al. 2010, 2011; see Hoang and Lazarian (2012) for a review) and observation (Dickinson et al. 2009; Kogut et al. 2011; Tibbs et al. 2012). In particular, *Planck* results have confirmed spinning dust emission as the most reliable source of the anomalous microwave emission (AME) (Planck

Collaboration 2011). *Planck* is poised to release important results on the CMB polarization; however, it remains unclear to what extent the spinning dust emission contaminates the polarized CMB signal. This has called for the revisiting of the problem of dust alignment via "resonance paramagnetic relaxation". Such studies were presented in Hoang et al. (2013, 2014).

Paramagnetic alignment of grains is inevitable and it is definitely present for interstellar grains. With normal paramagnetic materials and the accepted values of interstellar magnetic fields, the paramagnetic alignment is subdominant compared to RATs for the grains considered in **polarimetric** studies. However, for sufficiently small grains (Fig. 4b), RATs are negligible and one expects to see the signatures of paramagnetic alignment. Below, we show that the existing observational data is consistent with this conclusion.

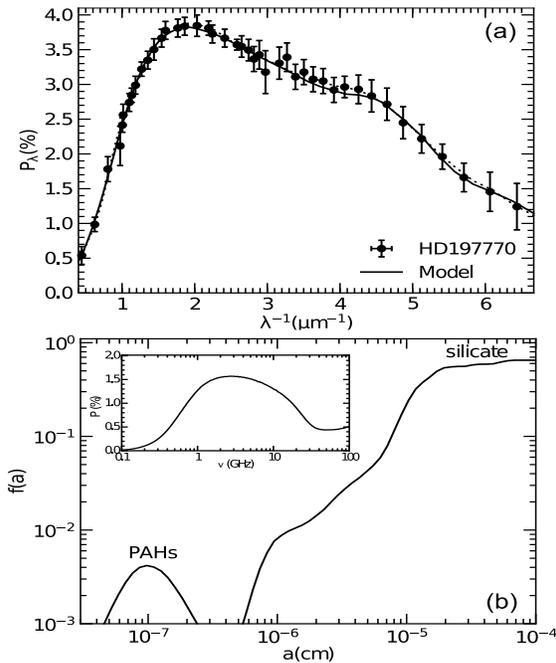

**Figure 12**. (a) Polarization curve for the best-fit model versus the observed polarization data for HD197770 from Wolff et al. (1997). The 3σ error bars are shown. (b) Alignment functions of silicate grains and PAHs for the best-fit model. Inset shows the polarization spectrum predicted for spinning dust emission. From Hoang et al. (2013). Reproduced by the permission of the AAS.

Since the PAHs that emit the spinning dust emission are likely to be the same grains that produce the UV extinction bump at 217.5 nm (see e.g., Draine and Li 2007), one can look for the signature of aligned PAHs using the starlight polarization at this wavelength.

The polarization bump at 217.5 nm toward two stars, HD197770 and HD147933-4, was discovered a long time ago (Clayton et al. 1992; Wolff et al. 1993), and it was suggested that the bump originated from aligned, small graphite grains (e.g., Draine 1989). Fitting to the observational data for these stars, Hoang et al. (2013) obtained the best-fit grain size distributions and alignment functions. For HD197770, it was found that the alignment of silicate grains alone cannot reproduce the UV polarization bump adequately. Instead, a model with aligned silicate grains plus weakly aligned PAHs can successfully reproduce the UV polarization bump as well as the polarization plateau. The inferred degree of PAH alignment present in HD197770 varies with the grain size and has a peak of 0.005 at ~ 0.9 nm. Although the degree of PAH alignment is rather small, due to the dominance of PAHs for particles with a<20 nm, it is sufficient to reproduce the UV polarization excess (Fig. 12). In fact, the low degree of alignment for PAHs is not unexpected from theoretical predictions based on resonance paramagnetic alignment, which was proposed by Lazarian and Draine (2000) and numerically studied in Hoang et al. (2014).

Due to the weak alignment of PAHs with the magnetic field, the spinning dust emission is weakly polarized. Hoang et al. (2013) found that the upper limit for the polarization of spinning dust emission is about 1.6% along the line of sight toward HD197770 at a frequency of 3 GHz (see the inset in Fig 12(b)). The derived constraint is consistent with a number of observational studies (Mason et al. 2009; Dickinson et al. 2009; Lopez-Caraballo et al. 2011; Macellari et al. 2011).

# 8 Observational tests of interstellar grain alignment

Hiltner (1949), the discoverer of interstellar polarization, already associated its origin with asymmetric grains aligned with the Galactic magnetic field. The association with dust was then borne out theoretically and confirmed observationally by comparing the amount of polarization with the visual extinction (e.g. Serkowski et al. 1975), which showed that the upper envelope of the polarization was linearly correlated with the color excess. The role of magnetic fields was confirmed by comparing the position angles of the optical polarization with those from the theoretically well understood mechanism of synchrotron emission, in objects where the two could be reasonably assumed to originate in the same location (e.g. Wolstencroft 1987; see also Andersson 2013). However, the fact that the amount of polarization is not a well-defined injective function of the extinction shows that the interpretation of observed polarization in terms of the column density of aligned grains is not a simple one.

As we discussed above, three theoretical paradigms have been put forward to explain the grain alignment: Paramagnetic relaxation, mechanical alignment and RAT. Numerous specific instantiations have been executed for each (see Section 6.1). While mechanical alignment -generally predicts polarization perpendicular to the projected field, in disagreement with observations, paramagnetic and radiative alignment predict polarization parallel to the field. A number of observations are available to probe the responsible alignment mechanism and include: the polarization for a given total visual extinction ($P/A_V$) in various environments; the wavelength dependence of the polarization, both broad-band and in specific lines (e.g. the aliphatic CH features); the dependence of the polarization on the angle between magnetic and radiation fields; and its dependence on the gas and dust temperature and chemical reaction rates (specifically the $H_2$ formation rate).

A number of observational issues need to be accounted for when interpreting the polarization in terms of grain alignment:

1) Dichroic extinction (or emission) by aligned grains only probes the plane of the sky component of the magnetic field.

2) We must have a population of grains with the physical properties to allow alignment torques to act on them (whether internal - as for paramagnetic grains; or structural - as for grains with a net helicity).

3) They must be sufficiently asymmetric to provide a net differential extinction.

4) Because polarization is a (quasi-) vector entity and because the polarization opacity is relatively small, line of sight averages can be complicated to invert into physical parameters.

5) Environmental changes affecting the grains and their surroundings can be difficult to determine accurately.

Hence, extracting information on the grain alignment mechanisms require careful observational designs and often a large amount of polarimetric and supporting data.

*8.1 Observational methodology*

The most fundamental measure of the fraction of aligned grains on a line of sight is the fractional polarization $P/A_V$. For a uniform magnetic field and constant grain alignment efficiency (and environmental parameters), this observable should be constant, and its value would, using assumptions about the grain size distribution and shape, provide information on the fraction of aligned grains.

The fractional polarization has the advantage that it can be employed at different wavelengths depending on what material is probed, with optical data yielding higher accuracy for diffuse gas and near infrared (NIR) data allowing studies of high opacity lines of sight (e.g. Jones et al. 1992, Gerakines et al., 1995, Arce et al. 1998). However, as shown by Myers and Goodman (1991), Jones et al. (1992) and discussed in a number of papers, e.g., by Ostriker et al. (2001), the topology of the magnetic field (including its turbulence) can cause the observed fractional polarization to depend sensitively on the line of sight structure of the field (see Fig.10b). As discussed by Jones et al. (1992) (see also Ostriker et al. 2001), turbulence alone can cause the fractional polarization to fall as steeply as $A_V^{-0.5}$ even with no inherent loss in grain alignment efficiency. This fall-off rate corresponds to a random walk process, and hence a significantly steeper slope would indicate unequivocal evidence for the loss of grain alignment.

By combining the fall-off in $P/A_V$ as a function of $A_V$, with mapping of the position angle distribution over the cloud surface, and assuming an isotropic velocity field, constraints can be put on the turbulent spectrum in the material (Jones et al. 1992; Andersson and Potter 2005). Together with theoretical modeling, some constraints can then be set on variations in the grain alignment efficiency along the line of sight (Whittet et al. 2008). Similar concerns apply to the far-infrared (FIR) and submillimeter (sub-mm) wavebands where we observe dichroic emission polarization (Ostriker et al. 2001). Based on single-band (or limited spectral coverage) observations, interpretation of the fractional polarization is limited not only by the uncertainties in the line of sight field topology, but also by possible variations in grain temperature, emissivity and other parameters in different parts of the cloud.

The characteristic wavelength dependence of the optical/near infrared polarization curve was first established by Behr (1959) and Gehrels (1960) and characterized by Serkowski (1973), who found that it could be well described by the functional form:

$$P = P_{max} \exp\left[-K \ln^2\left(\lambda_{max}/\lambda\right)\right] \quad (2)$$

where $P_{max}$ is the peak polarization at the wavelength $\lambda_{max}$ and $K$ describes the width of the curve. In Serkowski's original parameterization, $K$ was set to the fixed value 1.15, while subsequent work (Codina-Landaberry and Magalhaes 1976; Wilking et al. 1980) showed that the best-fit value varied from source to source. Wilking et al. (1982) and Whittet et al. (1992) argued that $\lambda_{max}$ and $K$ likely correlate and thus are not independent parameters.

As suggested by Codina-Landaberry and Magalhaes (1976) and shown through modeling by, e.g., Mathis (1986) the $K$-parameter is related to the average size of the aligned dust grain. Because several of the alignment mechanisms depend on the grain size and because the collisional disalignment is faster for smaller grains (Whittet 2003), the size distribution of aligned grains - as probed by the polarization curve (Kim and Martin 1995) - can therefore also be used to constrain changes in the alignment efficiency. Mathis (1986) attributed the cut off for the grain sizes as arising from the low probability of grains having a ferromagnetic inclusion - specifically the size where a single ferromagnetic inclusion was incorporated in the paramagnetic grain bulk. In the RAT theory, the cut-off arises from the drop of the efficiency of RATs as the grain size gets much smaller than the wavelength (see Fig. 4b).

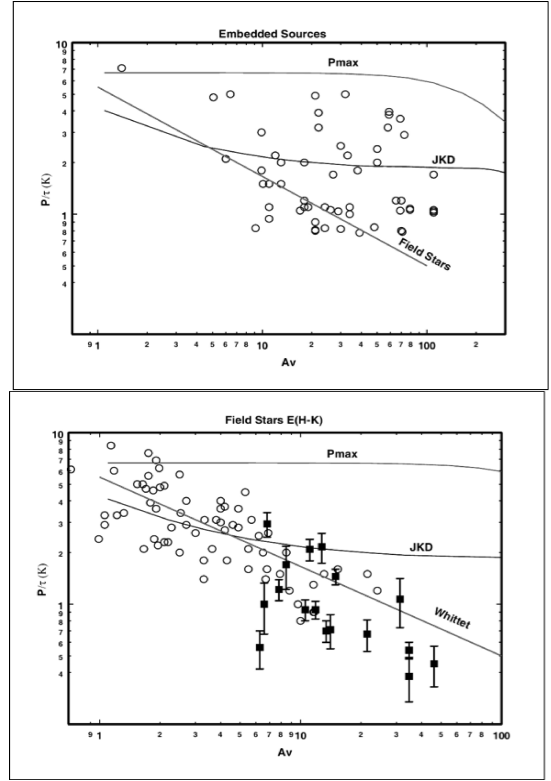

**Figure 13.** The fractional polarization in the K-band seen towards embedded sources (upper) and background field (lower) stars (data from Whittet et al. 2008) are compared with the predictions from a model of polarimetric radiative transfer through a cloud with a turbulent magnetic field

(courtesy T. J. Jones; for details see Jones et al. (2011)). Only at very large opacities without embedded sources does the fractional polarization drop below what can be explained by depolarization due to the random field component. See also Fig. 20.

Multi-wavelength polarization data overcome some of the limitations of single-band measurements of the fractional polarization, as shown by Andersson and Potter (2007) for dichroic extinction polarimetry and by Vaillancourt et al. (2008) for dust emission polarization. Since the line of sight effects of the magnetic field topology are wavelength independent, changes in the broadband spectral dependence of the polarization can be used to decouple grain alignment variations from the depolarization due to crossed field lines (and hence inherent depolarization) along the line of sight.

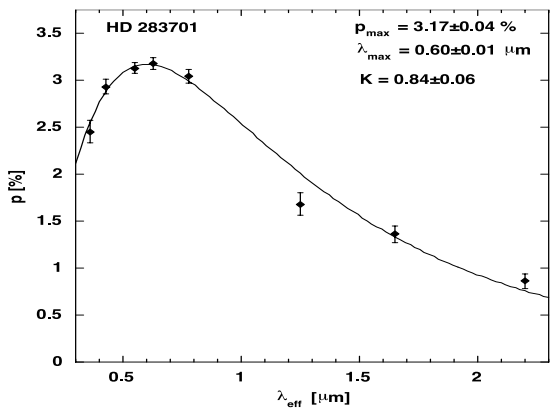

**Figure 14.** The wavelength dependence of interstellar optical/near infrared polarization follows a universal Serkowski relation (Eq. 2), here illustrated for the star HD 283701 behind the Taurus cloud. The data are from Whittet et al. (1992).

Polarization in discreet spectral line can also be a very powerful tool for evaluating alignment mechanisms and efficiencies, since it allows us to isolate either specific components of the dust, or specific locations within the clouds. For instance, strong polarization is seen in the 9.7 μm and 18 μm silicate spectral features (Smith et al 2000), but not in the 3.4 μm aliphatic C-H stretch feature, usually associated with carbonaceous dust (Chiar et al. 2006). Therefore, it is often assumed that large carbonaceous grains are either absent, symmetrical or unaligned. This is supported by the comparisons of linear and circular polarization in the ISM, which indicates that the dust grains giving rise to the polarization are good dielectrics (Martin 1974, Martin and Angel 1976, Mathis 1986), consistent with silicate, but not with carbonaceous grains. This should, however, be considered in the context of comprehensive dust models, based on inversions of the extinction curve and abundance constraints (see Kim and Martin 2995, Clayton et al. 2003, Draine and Li 2007, Hoang et al. 2014), which often require large amorphous carbon grains to be present. In RAT alignment the dichotomy between alignment of silicate and carbonaceous grains can be easily understood as due to the differences in the magnetic properties of the grains. While silicates are generally paramagnetic, carbon generally forms diamagnetic solids. Hence, even if spun up by radiative torques, the Barnett effect can not generate an internal magnetization for the carbonaceous grains, and they will therefore not align with the magnetic field.

Finally, the position angle of the polarization can also tell us something about the alignment mechanism. For both paramagnetic relaxation alignment and RAT alignment, the grains are expected to be oriented with their long axes perpendicular to the magnetic field lines, yielding a polarization parallel to the projected field. In contrast, in the standard interpretation of mechanical ("Gold") alignment the grains have their long axes oriented parallel to the field lines (Gold 1952a,b; Lazarian 2007). This occurs because the grains are aligned by the relative motion between the gas and the dust (tumbling like water-wheels in a stream) provided that we can assume, as is usually done, that the partially ionized gas is restricted to flowing along the magnetic field lines.

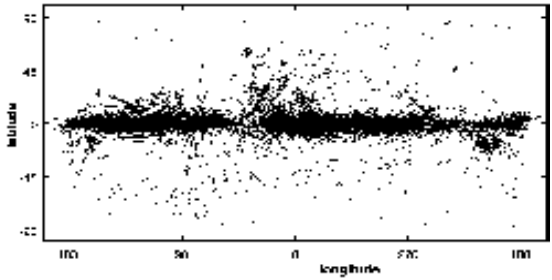

**Figure 15.** Large scale mapping of optical polarimetry towards stars in the Galaxy shows an ordered field with the main part of the polarization vectors oriented in the plane of the Galaxy. Together with studies of synchrotron radiation and Faraday rotation, this implies that the Galactic field is ordered and - predominantly oriented in the Galactic plan and that the position angles of optical/NIR polarization are parallel to the magnetic field. Data are from Heiles (2000).

On global scales, Faraday rotation observations show that the Galactic magnetic field is oriented in the plane of the disk (e.g., van Eck et al. 2011 and references therein), so mechanical alignment predicts optical polarization with position angles predominantly perpendicular to the Galactic plane. This is contrary to observations, placing severe doubt on Gold alignment as a viable general alignment mechanism.

Turbulence dominated by Alfven waves can induce Gold alignment perpendicular to magnetic field (Lazarian 1994, 1995). However, the random mechanical torques responsible for the Godl alignment are in most cases subdominant compared to RATs (see Lazarian 2007).

*8.2 Observational considerations*

In addition to the complications arising from the line of sight integration of the polarization discussed above, a number of observational effects need to be considered when interpreting interstellar polarimetry in terms of grain alignment. Primarily, in addition to considering the alignment drivers, the observed polarization will also depend on the mechanism and strength of the disalignment processes (i.e., grain orientation randomization). Draine and Lazarian (1998) reviewed the theoretical disalignment mechanisms and argued that, for neutral interstellar material (which most dust polarization observations probe), gas-grain collisions should dominate the grain randomization. Since the collision rate depends on both the gas temperature and density - and other disaligning mechanisms depend on the grain temperatures, as well as the gas and dust ionization states - an understanding of these parameters is also needed to fully constrain the alignment physics. While little direct observational data exist on grain disalignment (e.g., Andersson and Potter 2007), we have recently argued (Andersson et al. 2013) that the depolarizing effects of a strongly enhanced collisional rate may have been detected in the compression ridge of the reflection nebula IC 63.

Thermodynamic arguments (Jones and Spitzer 1967) require that, to maintain an anisotropic distribution of the grain spin-axes, a "heat engine" is active between the mechanisms driving and damping the alignment. Under the assumption that gas-grain collisions dampen the alignment, this means that for alignment mechanisms dependent on the internal grain dynamics, e.g. paramagnetic alignment, the gas and dust temperatures much be significantly different for alignment to be efficient.

If radiative processes drive the grain alignment, it is particularly important to note that the line-of-sight extinction towards a background source ($A_V$), used to calculate the fractional polarization, or to evaluate the characteristic depth into the cloud of a given parcel of dust, is not necessarily a good indicator of the amount of radiation incident upon the grains causing the polarization. The use of the line of sight extinction for these purposes can cause significant outliers in correlations between tracers of the grain alignment and column density. Andersson and Potter (2007) used a comparison of the grain alignment (as traced by the location of $\lambda_{max}$) and the 60 μm /100 μm color temperature, as functions of the line of sight extinction to show that for highly asymmetric clouds, or clouds close to localized radiation sources, the polarization can significantly differ from the average for similar surrounding regions. A similar effect is seen in the data from Whittet et al. (2008) where lines of sight close to, or towards, embedded sources show, on average, an elevated fractional polarization compared with those towards background field stars not passing close to embedded young stellar objects (YSOs).

Because the radiation environment and average grain size distributions (as well as other likely parameters, including grain composition) vary from cloud to cloud, it is important to carefully consider if, or how, to combine observational data acquired in different regions. Andersson and Potter (2007) showed that a general $\lambda_{max}$ vs. $A_V$ relationship exists for six local clouds (such a correlation was first noted, for the Taurus cloud, by Whittet et al 2001). However, the zero intercept in this relation $\lambda_{max}$ ($A_V=0$) differs from cloud to cloud and is proportional to the average total-to-selective extinction ($R_V$) of the clouds. They interpreted this as a result of slightly different grain size distributions in the clouds. For each individual cloud no correlation between $\lambda_{max}$ and $R_V$ was seen, whereas if the data (from all clouds) were combined, a relationship consistent with that of Whittet and van Breda (1978) was recovered.

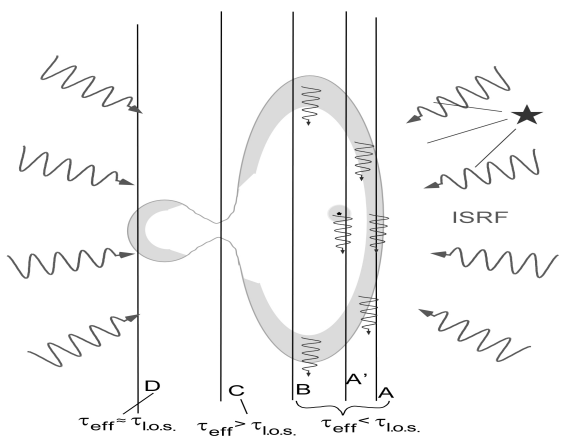

**Figure 16.** The line of sight opacity is not always a good tracer of the radiation field impinging on a dust. For non-spherical clouds or if there are stars near by the cloud the use of $A_V$ as indicator of the relative illumination of the dust can be misleading. From Andersson and Potter 2007; reproduced by the permission of the AAS.

For dust emission polarimetry, several of the above caveats also apply, including effects of proximity to illuminating sources. In addition, the emission from the dust depends not only on the grain size distribution, but also on their temperature and emissivity. The grain temperature at the column densities required for current FIR and sub-mm wave observations are likely to have a narrow range (at least for starless clouds, c.f., Hollenbach et al. 2009). However, in deeper parts of the clouds, where grain growth is likely to occur, both the size distribution and emissivity of the grains may change. While instrumental capabilities have been dramatically improved over recent years, it is still difficult to compare sub-mm wave polarimetry at different wavelengths due to the often very different angular scale sizes. In contrast to the pencil beam measurements of extinction polarimetry, the [differing] beam averages inherent in emission observations require great care in interpreting multi-wavelength data in terms of physical processes in *a given* dust component. For example, data from experiments designed to study the cosmic microwave background (e.g., *WMAP*, *Planck*) often measure dust polarization at wavelengths of 1 to 3 mm, but with much lower spatial resolution (~ 0.3 to 1°) than measurements at shorter wavelengths (~ 5--40 arcsec). While the FIR and sub-mm data sample dense dusty regions ($A_V$ ~ 20 - 40 mag) the CMB data is dominated by diffuse emission or has equal parts diffuse and dense emission (Bierman et al. 2011).

*8.3 Observational tests of grain alignment theories*

We first note that optical and infrared observations probe the "classical" dust grains with approximately $a \geq 0.01$mm, and we will limit our discussion to this grain population here.

**Paramagnetic relaxation alignment**

Because of the limited direct information available of the grain mineralogy, the observational prediction of paramagnetic relaxation alignment can be difficult to turn into unequivocal observational tests. The classical Davis-Greenstein mechanism assumes that the grains are paramagnetic---consistent with silicate grains---and spun-up by gas-grain collisions to rotational energies corresponding to the thermal energy of the gas. Jones and Spitzer (1967) used fundamental thermodynamic arguments to show that for such thermally spinning grains to remain aligned, the magnetic susceptibility must significantly exceed that of typical paramagnetic materials.

The magnetic properties of interstellar grains are difficult to ascertain using remote sensing. Interstellar depletion measurements (Jenkins 2009) together with dust size distributions based on the extinction curve, place constraints on the composition of grains (Mathis, Rumpl, Nordsieck 1977; Kim and Marin 1995; Clayton et al. 2003), but not on the mineralogy. Near- to mid-infrared spectroscopy can yield some information about the grain mineralogy (van Dishoeck 2004, and references therein) and interplanetary dust particles collected in the Earth's upper atmosphere provide some direct experimental constraints. Even so, conclusively determining the magnetic susceptibility of interstellar grains is likely not possible.

Jones and Spitzer (1967) suggested that the required enhancement of the grain susceptibility could be accomplished, without violating elemental abundance constraints in the solid phase of the ISM, by including small ferromagnetic sub-grains into the silicate grain bulk. Mathis (1986) expanded on this suggestion and showed that the observed polarization curve could be reproduced using the grain size distribution derived by Mathis et al.(1977) and the simple assumption that a silicate grain is aligned if, and only if, it contains at least one such super-paramagnetic (SPM) inclusion. The identification of amorphous silicate grains in interplanetary dust particles (Bradley 1994) seemed to support the hypothesis. Goodman and Whittet (1995) argued that the dark Fe (Ni)-metals and iron-rich sulfide patches seen inside these grains were of the right size and volume-filling factor to match the requirements of Mathis' theory and exhibit super-paramagnetic susceptibilities.

However, using a large database of polarization and depletion measurements, Voshchinnikov et al. (2012) showed that polarization efficiency is likely not associated with super-paramagnetic grain materials. In their study, they measured the interstellar depletion in silicon, magnesium and iron. Assuming that the silicon is fully incorporated into grain materials with the stoichiometry of olivine, they could then calculate the iron that could be in ferromagnetic solid form. No statistically significant correlation was seen when comparing the fractional polarization with the fraction of ferromagnetic solid iron.

Another observational argument against paramagnetic relaxation alignment is based on the thermodynamic argument of Jones and Spitzer (1967), strengthened by Roberge (1996), that for (super) paramagnetic alignment to be possible, the gas and dust temperatures must be significantly different (see quantitative estimates in Roberge and Lazarian 1999). Jones et al. (1984) used H-band polarimetry of Tapia's Globule #2 in the Southern Coalsack, supported by FIR photometry and CO (J=1-0) line emission to test this prediction. They showed that although the estimated dust and gas temperatures were almost identical, at ~10 K, significant polarization was observed toward the cloud core, indicating at least partial grain alignment. Because these temperature (and polarization) determinations are dependent on line of sight averages and possible radiative transfer effects, these results were, however, not conclusive in terms of the viability of paramagnetic relaxation alignment.

However, polarization in the CO ice line at 4.7 μm towards W33A (Chrysostomou et al. 1996) and, particularly, towards the background star Elias 16 (Hough et al. 2008) with $A_V$ =21.6 mag (Murakawa et al. 2000), behind the Taurus cloud, conclusively shows that grains deep into the clouds are aligned. CO ice is destroyed at visual extinctions less than ~6 mag (Whittet 1989) and rises in relative abundance to reach about 50% only at $A_V$ =20 mag (Whittet et al. 2010). Hence polarization in the interstellar CO ice feature shows that the grains are aligned at least to $A_V$=6 mag and probably significantly deeper. It is worth noting here that the polarization spectrum for Elias 16 (Hough et al. 1988) with $\lambda_{max}$=1.08±0.08 µm is fully consistent with the $\lambda_{max}$ vs. $A_V$ relation seen by Andersson and Potter (2007) based on low- to medium-opacity data. Not only does the existence of CO ice itself put direct limitations on the temperature of the gas in these regions, models (e.g., Hollenbach et al. 2009) show that the difference in gas and dust temperatures at these opacities should approach zero.

Finally, Andersson et al. (2013) suggested that $H_2$ formation yields an enhancement in the polarization seen in the photo-dissociation region (PDR) around IC 63. If RAT alignment is correct, this would indicate that the grains cannot be super-paramagnetic, since, as shown by Hoang and Lazarian (2009a) super-paramagnetic grains yield saturated alignment where any additional drivers (e.g., Purcell torques) would not be able to further enhance the alignment. A detailed modeling of grain alignment in IC 63 by Hoang et al. (2015) showed that H2 formation can enhance the dust polarization.

**Radiative Grain Alignment**

From the point of view of the observer, the fundamental prediction of RAT is that grain alignment will be efficient only if the grain is exposed to anisotropic radiation with a wavelength of $\lambda$ <2*a*, where *a* is the effective grain radius. In practical terms, the first condition is almost always fulfilled as, either the radiation emanates from a localized source, or the diffuse interstellar radiation field is attenuated by discreet clouds with a well-defined density gradient. Because both the spin-up torques and alignment torques (with the magnetic field) are driven by the radiation, RAT also provides the unique prediction that the alignment efficiency should depend on the angle between the radiation field anisotropy direction and the magnetic field orientation. The RAT theory that we discussed earlier in the review thus leads to a number of observational tests:

*1. Is the grain alignment enhanced at stronger radiation fields?*
*2. Do changes in the color of the radiation field lead to changes in the size distribution of aligned grains and thus changes to the polarization curve?*
*3. Does RAT alignment predict a size distribution of aligned grains consistent with that derived from extinction and polarimetry in different environments?*
*4. Does the alignment efficiency vary with the angle between the radiation anisotropy and the magnetic field?*

For all of these questions the answer is "yes" and we will now elaborate on the pertinent observations. It should first be noted that the lack of polarization in the CH aliphatic spectral feature at 3.4 micron requires a role for mineralogy in the alignment. Most inversions of the interstellar extinction curve (see, e.g., Hoang et al. 2014) require large carbonaceous grains. Since such grains, when exposed to the UV and cosmic ray field of the ISM, will develop a surface layer of aliphatic bonds (Chiar 2013, private communication), the lack of a polarization signal in the 3.4 mm line would imply that these large carbonaceous grains are either spherical, have no helicity or are not affected by the radiation. However, as mentioned above, the lack of a Barnett magnetization in diamagnetic solids, may simply mean that, while the carbon grains are spun up by RATs, the do not interact with the ambient magnetic field and thus do not align in the general ISM.

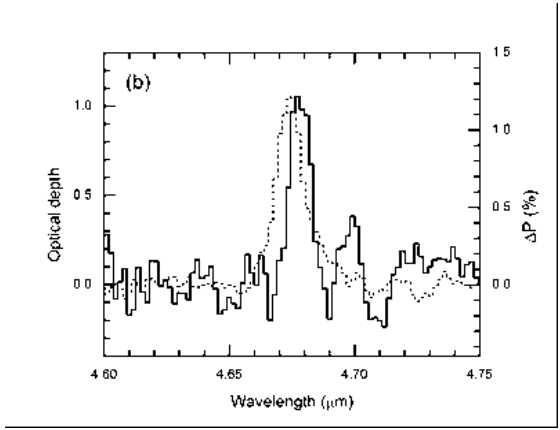

**Figure 17.** The CO ice line towards the background star Elias 16 shows significant polarization, indicating that grains are aligned at $A_V>6$ mag. The dashed line (left-hand scale) shows the opacity while the full line (right hand scale) shows the polarization, with the continuum removed (from Hough et al. 2008)

Whittet et al. (2008) used NIR polarimetry and RAT-based modeling to show that the drop in the fractional polarization over a wide range of visual extinctions is consistent with radiative alignment. However, they find a drop-off in $p/A_V \sim A_V^{(-0.52\pm0.07)}$ which is still consistent with turbulent magnetic fields. They also found that the normalization of the $p/A_V$ vs. $A_V$ relation indicates a better alignment efficiency for a line of sight terminating in an embedded source, than for those towards true background sources. This enhancement can be understood as being due to radiative alignment by the radiation from the Young Stellar Objects (YSO), but might be dependent on other possible dust and environmental changes induced by the YSOs.

More unequivocal support for enhanced alignment efficiency near localized sources comes from studies where lines of sight toward true background stars pass close to specific stars illuminating the dust. Andersson and Potter (2010) studied the polarization in the Chamaeleon I cloud close to the star HD 79300. This star has been shown to be located ~0.03 pc from the near cloud surface (Jones et al 1985; using an updated stellar luminosity from Luhman 2004). They find that the dust heating is dominated by the stellar radiation out to a projected distance of ~0.75° and a clear linear correlation exists between the alignment efficiency and the dust temperature towards the background stars. Matsumura et al. (2011) similarly find that the polarization efficiency is enhanced when the dust temperature is elevated for stars in the Pleiades cluster. Finally, Cashman and Clemens (2014), using imaging H-band polarimetry of a number of sub regions in LDN 204 find that the polarization efficiency for the different sub-regions is enhanced as the projected distance to $\zeta$ Oph decreases.

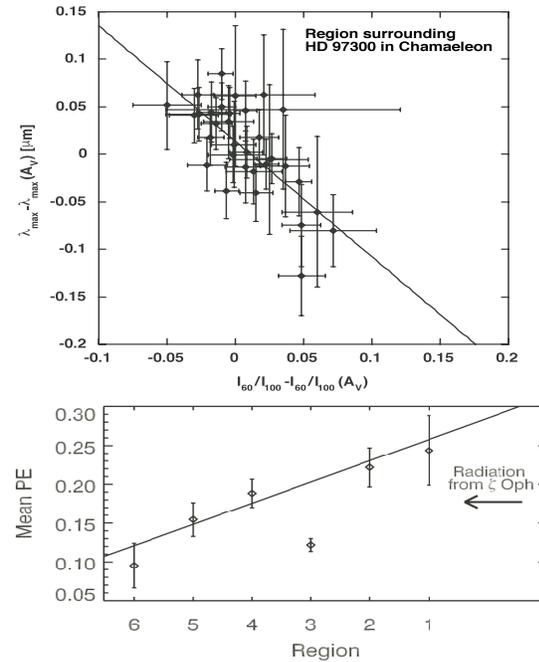

**Figure 18.** The alignment efficiency can be seen to be enhanced by the proximity of a bright star. Top: for the region around HD~97300 in Chamaeleon I, the alignment efficiency is enhanced (smaller $\lambda_{max}$ -$\lambda_{max}$ ($A_V$)) as the grain heating from the central star increases. Differential entities are used to account for the different path-length through the cloud towards the background sources (Andersson and Potter 2010; reproduced by the permission of the AAS). Bottom: the mean polarization efficiency for a number of region within LDN 204 show a systematic enhancement with lessening projected distance to the illuminating star $\zeta$ Oph (Cashman and Clemens 2014; submitted; reproduced by permission of AAS).

As shown by Kim and Martin (1995) and discussed above, the shape of the optical/NIR polarization curve can be inverted to provide information on the size distribution of aligned grains. If RAT theory is valid, changes in the aligning radiation field energy distribution should result in changes in the size distribution of aligned grains and hence in the polarization curve.

In their modeling, Kim and Martin (1995) inverted a representative set of interstellar polarization curves to predict the size distributions of aligned grains. They found that for polarization curves characteristic of the diffuse ISM the aligned grains had a small size cut-off around $a_{ali}$=0.04-0.05μm, much larger than the small grain size cut-off derived from extinction curve inversion. Given that neutral hydrogen in the ISM will remove all photons shortward of the Lyman limit (91.2 nm $\approx 2a_{ali}$) this indicates that the small size cut-off in aligned grains (for diffuse gas) might be simply understood as a direct result of RAT alignment requirements.

Because the interstellar extinction curve rises to the blue, the remaining light at larger and larger depth into a cloud becomes successively redder. Based on RAT theory this should mean that the small size cut-off in the distribution of aligned grains should grow and, therefore, the peak of the polarization curve should shift to longer wavelengths. This is seen (Andersson and Potter 2007) as a universal linear relation between $\lambda_{max}$ and $A_V$ for (at least) six local interstellar clouds.

Using an argument closely parallel to that of Mathis (1986), RAT alignment can also explain the correlation between $K$ and $\lambda_{max}$ seen in ISM polarization (Whittet et al. 1992). Namely, if the underlying grain size distribution is fixed and the smallest aligned grain is determined by the wavelength cut-off of the available radiation, then the polarization curve will narrow ($K$ increase) and shift to the red, with a reddened radiation field, as is observed.

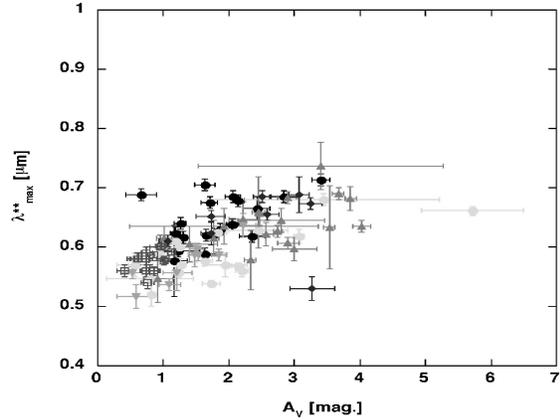

**Figure 19.** The wavelength of maximum polarization is seen to be linearly correlated with the visual extinction. Different symbol shades correspond to data for different near-by clouds and the raw data have been adjusted to account for different average grains sizes (via $R_V$) and star formation rates on the different clouds. See Andersson and Potter (2007) for details.

Because of constraints imposed by the total elemental abundances of refractory materials in the ISM, there must be an upper bound on the grain size distribution (Kim and Martin 1995). In very deep (starless) clouds, RAT theory implies that the alignment should cease once the diffuse radiation field has been reddened such that the shortest remaining wavelengths in the cloud are larger than the diameters of the largest grains. Whereas, as discussed above, the fractional polarization can decrease due to the magnetic field topology and turbulence, even without a decrease in the alignment efficiency. Jones et al. (1992) showed that in the fully turbulent limit the observed fractional polarization behaves according to a random walk process and shows a functional dependence of $A_V^{-0.5}$. Once no further polarization is added with increasing opacity $p/A_V$ should fall as $A_V^{-1}$. If we use the RAT condition for alignment, a standard ISM extinction curve (e.g., Cardelli et al. 1989) and the best estimates of the upper grain size distribution cut-off, we find that this loss of grain alignment should occur at $A_V$ ~10 - 25 mag., depending on the value of the total-to-selective extinction and upper grain size limit assumed. So far, dichroic extinction polarimetry has not been able to systematically probe beyond this depth (Whittet et al. 2008; Jones et al. 2011).

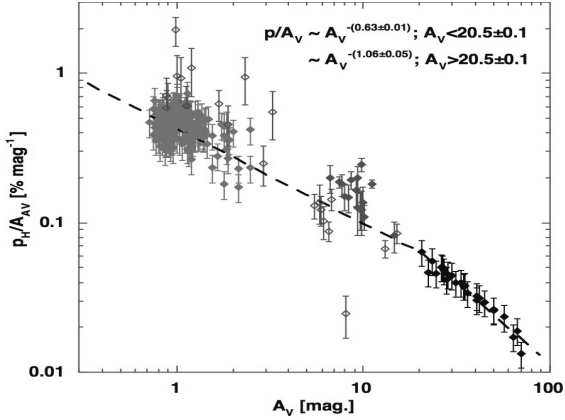

**Figure 20.** RAT alignment predicts that, once the reddened light at the center of a starless cloud has a wavelength larger than the largest grains, the grain alignment should cease. Combining optical, NIR and sub-mm wave data (black) we have, tentatively identified this break point in LDN~183, at $A_V$ ~ 20 mag.

However, as shown by Hildebrand (1988), emission polarization can be put on the scale of extinction polarization by multiplying with the dust opacity. The absolute scaling between emission and extinction polarimetry is uncertain due to the often poorly determined grain emissivity and temperature distributions. Combining optical, NIR and sub-mm wave polarimetry of the core of LDN 183 (allowing an arbitrary constant offset between NIR and sub-mm wave data) we have been able to, tentatively, locate a break in the fractional polarization at $A_V$=20 mag, below which a power-law index consistent with -0.5 is found and above which an index of -1 is found. This result is consistent with RAT theory predictions and would have important consequences for interpreting polarimetry towards highly extinct regions. Recent observations by Alves et al. (2014) and Jones et al. (2015) confirm these results, indicating that alignment is indeed generally lost at large opacities, of typically A_V=20 mag.

The FIR/sub-mm wave polarization spectrum for star forming cores shows a minimum around 350 µm. As shown by Hildebrand et al. (1999; c.f., Vaillancourt 2002) a single population of grains cannot produce this spectrum. At minimum, a two-component dust population is required with the two components possessing, either significantly different grain temperatures, or different wavelength-dependent emissivities. Using a model, where the components differ in temperature and the warmer dust component is better aligned, they could reproduce the observed drop in polarization from 60 to 300µm. In such a two-component model, the increase in the polarization spectrum long-ward of 350µm can also be understood as an effect of different grain emissivities (Vaillancourt et al. 2008; Draine and Fraisse 2009). Since the dust heating in star forming cores is expected to be dominated by the embedded YSOs, this strongly supports a radiative grain alignment mechanism. Further support for this interpretation, and expansion of the results, comes from Vaillancourt and Matthews (2012) and Zeng et al. (2013).

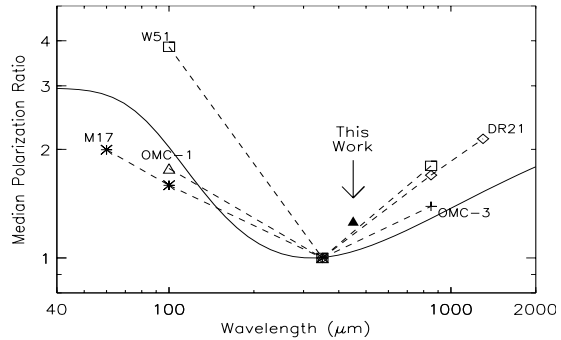

**Figure 21.** The observed polarization spectrum for a number of molecular cloud cores shows a systematic behavior with a marked minimum in the 350µm range. Overlaid (solid line) is an illustrative two-temperature dust model where only the hot component is polarized (from Vaillancourt et al. 2008, © AAS, reproduced with permission).

In the RAT paradigm, the torques aligning the grains with the magnetic field are also due to radiation. Since the torques act during the Larmor precession of the spinning grain around the field lines, the alignment efficiency with the magnetic field should depend on the angle between the radiation anisotropy vector and the magnetic field direction (Ψ; LH07a). Specifically, for the diffuse medium RAT theory predicts that grain alignment should be most efficient at Ψ=0.

Andersson et al. (2011) used multi-band polarimetry of the region around HD 97300 to probe this prediction. They find a $9\sigma$ deviation from a null result for the grain alignment as a function of $\Psi$ with the location of the maximum alignment efficiency consistent with $\Psi=0$. They also detect a sinusoidal variation in the 60µm /100µm color temperature with $\Psi$, which they model with a simple phenomenological model of "pizza box" like dust grains heated by the central source. Assuming the grain axis ratio from Kim and Martin (1995), they find both the variations in the grain alignment efficiency and 60µm /100µm with $\Psi$ are consistent within their simple model.

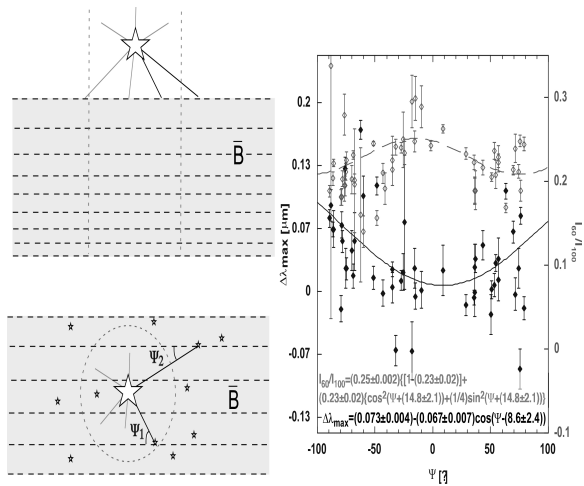

**Figure 22.** For a star close enough to an interstellar cloud to dominate the radiation field, RAT theory predicts that the alignment should vary with the angle between the magnetic and radiation fields. We can probe this by observing background targets located around the illuminating star (left). In accordance with predictions, Andersson et al. (2011) found a maximum in the grain alignment (minimum in $\Delta\lambda_{max}$) at $\Psi=0$ (filled symbols). The open symbols and dashed curve show the color temperature and best fit to a toy model of the differential heating of the aligned grains (see Andersson et al. 2011 for detail; reproduced with permission © ESO ).

**Purcell ($H_2$ formation) torque alignment**

Classical "Purcell alignment" - paramagnetic relaxation alignment of grains, driven to high spin rates by the ejection of newly formed $H_2$ molecules (or photo-electrons) from their surfaces (Purcell 1979), is unlikely to work due to the thermal trapping mechanism discussed by Lazarian and Draine (1999a,b). However, grain spin-up due to $H_2$ formation and ejection is predicted to play a role also in RAT alignment (Hoang and Lazarian 2009a). We recently showed (Andersson et al. 2013) that, in the PDR of IC 63, a correlation between the intensity of the $H_2$ 1-0 S(1) line at 2.122 µm and the polarization can be seen. Because the evolution of the PDR is slow on the time scales of the [$H_2$] chemistry (Morata and Herbst 2008), and the dissociation of molecular hydrogen proceeds via a two-stage process initiated via a line excitation into bound electronic states, the fluorescent intensity on the NIR line is directly proportional to the formation rate of $H_2$. This implies a role for "Purcell-rocket torques" in grain alignment. While more, and higher signal-to-noise (S/N), data are required to confirm these results, the dependence of the polarization on $H_2$ fluorescent intensity is consistent with our model of RAT alignment in a PDR.

In addition, detailed *ab initio* modeling of the polarization curves for regions with and without high $H_2$ formation rates in a RAT alignment paradigm yield very good fits to the red part of the optical polarization--based on reasonable dust characteristics. Short-ward of ~450nm, the models predict a measurable, relative enhancement in the polarization for regions with high $H_2$ formation rates. At these shortest wavelengths, the data in Andersson et al. (2013) are not quite high enough in S/N to be able to test this prediction, but the predicted effect is large enough to be well within reach of modern polarimeters on large telescopes.

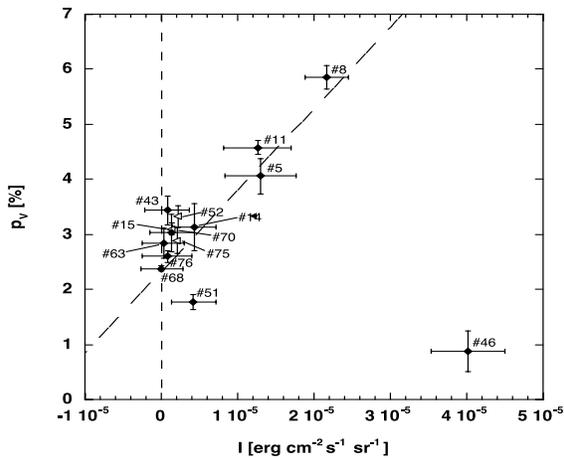

**Figure 23.** The amount of polarization towards stars behind the reflection nebula IC 63 is correlated with the fluorescent intensity in the $H_2$ 1-0 S(1) line. The fluorescence traces the $H_2$ destruction but also the $H_2$ formation because the time scales for reformation of $H_2$ on the dust grain surfaces is much shorter than the dynamical time scale of the PDR. Target #46 is located behind the compression ridge of the nebula where intense collisional disalignment dominates. Reproduced from Andersson et al. (2013) by the permission of AAS.

*8.4 Why aren't the carbonaceous grains aligned?*

As discussed above, comprehensive models of the dust distribution (Clayton et al. 2003; Draine and Li 2007) tend to require a component of large amorphous carbon grains. During their processing in the winds of red [asymptotic] giant stars and the ISM, such grains are expected to develop a surface coating of aliphatic C-H groups (Chiar 2013; private communications). Spectral line polarimetry of the 3.4 μm aliphatic C-H stretch feature (Chiar et al. 2006) has, however, not detected any polarization from such grains. Additionally, as noted above the comparisons of linear and circular polarization in the ISM, indicates that the dust grains are good dielectrics (Martin 1974, Martin and Angel 1976, Mathis 1986), consistent with silicate, but not with carbonaceous grains. Does this mean that all carbonaceous grains are symmetrical? Or, that they possess no helicity? Or that the comprehensive dust models are wrong? Or, that something is still missing in RAT theory?

While further observational work is needed to address these questions, our ongoing study indicated that indicates that the low efficiency of the alignment of carbonaceous grains might be attributed to their much smaller magnetic moment. Indeed, the main process for the grains to acquire a magnetic moment is related to the Barnett effect (Dolginov and Mytrophanov 1976), which acts in paramagnetic materials. While carbonaceous grains are likely to have unpaired electrons, their paramagnetic response is expected to be weaker than for silicates (particularly for a disordered carbon grain). As a result, the grain precession rate gets much slower and may become comparable with the RAT alignment rate. In this case, the alignment does not happen in respect to magnetic field but in respect to radiation direction with variation in the direction of anisotropy, including cancelation of polarization.

In addition, our DDSCAT calculations indicate that the RATs acting on carbonaceous grains are an order of magnitude smaller compared with the torques on silicate grains. Naturally, these ideas require further studies, in particular more detailed studies of grain randomization, including the randomization from electric dipole fluctuations (Jordan and Weingartner 2009), which are important for grains moving with high velocities produced by dust interactions with the ambient MHD turbulence (Lazarian and Yan 2002, Yan and Lazarian 2003, Yan et al. 2004, Hoang, et al. 2012).

While the issues of the alignment of carbonaceous grains are studied further, it should be noted that the (presumptive) lack of alignment of large carbonaceous grains is *not,* in and of itself, an argument *in favor* of paramagnetic alignment, since the magnetic susceptibility of graphite (in plane) is within an order of magnitude of e.g. olivine (Belley et al. 2009).

**9 Future work**

While a substantial progress has been achieved in the field of grain alignment and its testing, there are still important issues to address, we feel. The observed low degree of the alignment of carbonaceous grains is still requires more theoretical/observations study.

Another issue is the role of the mechanism of alignment with RATs and enhanced paramagnetic (superparamagnetic, ferromagnetic, ferrimagnetic…) relaxation working hand by hand as described in Lazarian & Hoang (2008). In that work we explained that the enhanced relaxation should stabilize the high-J attractor point resulting in the perfect 100% alignment. This can explain the cases when the alignment is higher than it is expected on the basis of the pure RAT theory. This stabilization is also important in the case if the processes of randomization are locally enhanced. Note, that in view of high abundance of iron in interstellar media the presence of inclusions with higher response is not unexpected. Nevertheless, we show in Hoang & Lazarian (ApJ, submitted) that thermally rotating grains with strongly magnetic inclusions are weakly aligned only.

The alignment of large grains does require theoretical work as well. As the probability of having strongly magnetic inclusions increases with the grain size, we can show that with strongly magnetic response we can explain perfect alignment of large grains that cannot be well aligned only by RATs. In addition to stabilizing of high-J attractor points, or large grains the presence of inclusions increases the Larmor precession, that we have shown to be an essential part of the RAT mechanism.

**10 Discussion**

Grain alignment has been theoretically studied for decades but the solution seemed to be elusive. It is only recently that both theoretical progress and advances in observational study of astrophysical polarization brought us to the stage when grain alignment can be considered a predictive theory. This theory is essential for relating the observed polarization and underlying magnetic fields.

The most important development of the recent years is the identification of radiative torques as the major agent of alignment for grains that are responsible for both emission and extinction polarization employed for studying astrophysical magnetic fields. With helicity identified as the cause for RAT alignment and the analytical model describing this alignment, it became possible to quantify grain alignment and provide quantitative predictions for the polarization from various astrophysical environments, not only diffuse interstellar medium and molecular clouds, but also from accretion disks, comets, and circumstellar regions.

Traditionally grain alignment theory assumed the grains to be either prolate or oblate ellipsoids. Such grains do not have helicity. Therefore to provide helicity to the model grain in the simplest manner, LH07a attached a mirror to the oblate body of the grain and showed a remarkable similarity between the torques acting on the adopted toy model of a helical grain with the properties of the torques on actual irregular grains calculated with the DDSCAT code.

The simplicity of the predictions of the LH07a, and subsequent studies, allow easy parameterization of the RAT alignment within numerical codes. This allows the theory to provide reliable quantitative polarization predictions. The current work in this direction usually adopts *ad hoc* ideas about grain alignment rather than a quantitative theory. Proper modeling of alignment should substantially increase the scientific output of the current and future polarimetry observations.

This review is devoted to the recent developments in the theory of RAT alignment and its observational testing. However, it also deals with the theoretical understanding of the paramagnetic alignment of very small grains. We show that the mechanisms are complementary. For instance, RATs are inefficient in aligning grains for which $a \ll \lambda$ (Fig. 3b). Therefore for sufficiently small grains other mechanisms should be important. For instance, Lazarian and Draine (2000) introduced a new process, "resonance paramagnetic relaxation," and showed that it can be important for the alignment of very small grains or PAHs, which are responsible for the galactic anomalous microwave emission (Draine and Lazarian 1998a,b; Hoang et al. 2010, 2011). The alignment of PAHs at a level of 1% is quantified in Hoang et al. (2013) using the 217.5 nm polarization feature. There can be circumstances when other alignment processes take over. It is the goal of the future research to identify them, on the basis of comparing theoretical predictions and observations.

We would like to stress two new recent developments related to grain alignment theory. First, the theory has become quantitative and now it is able to account for the existing observational data. Another important development is that the grain alignment is no longer a concern of only interstellar experts. Examples in this review as well as in Lazarian (2007) convincingly show that one should account for aligned grains in many diverse environments, e.g. accretion disks, circumstellar regions, interplanetary space and comets (see also Hoang and Lazarian 2014). Understanding the polarization from dust in the diffuse media is also a burning issue for CMB polarization research. We believe that combining emission and extinction polarization it is possible to address many scientific issues that cannot be investigated by just one technique.

RAT alignment has proven to be the dominant mechanism for grains in many environments. For some environments, more research into the details of RAT alignment is necessary. For instance, the study of the alignment of grains in the absence of internal relaxation (Hoang and Lazarian 2009b) is not capable of predicting the grain alignment degree with the accuracy available for the alignment of grains when the internal relaxation is important (Lazarian and Hoang 2007a, 2008; Hoang and Lazarian 2008, 2009a, b). This results in more uncertainties for modeling polarization from accretion disks, where large grains are known to exist. Additional work is also necessary to explain why carbonaceous grains are far less aligned than silicate one.

As it is not possible to cover the subject of grain alignment within a short review, we refer our readers to earlier reviews, where particular facets of the grain alignment problem are presented in more detail. The most detailed review to this moment is Lazarian (2007), which, however, does not cover the most recent developments. Thus Lazarian (2007) is complementary to the present review. Reviews written earlier than 2007 are mostly outdated in terms of the RAT alignment. However, the Lazarian (2003) review may be recommended for those who are interested in the history of grain alignment ideas. The connection between grain alignment and CMB polarization studies is discussed in the Lazarian (2008) review. Aspects of the observational problem are addressed in Andersson (2012). It is important that the theoretical predictions are being successfully tested observationally, as it is discussed in the review.

**Acknowledgements** We thank John Vaillancourt and anonymous referees for many helpful suggestions. We acknowledge support: A.L. and B-G A. from NSF Grant AST-1109469, A.L. from NASA Grant NNX11AD32G, Vilas Associate Award and the NSF Center for Magnetic Self-Organization. A.L. and T.H. thank the Natal International Institute for Physics for hospitality. T.H. was supported by Humboldt Fellowship at Ruhr University in Bochum and Goethe University in Frankfurt, Germany.